\documentclass[10pt,manuscript,nonacm]{acmart}

\AtBeginDocument{%
  \providecommand\BibTeX{{%
    \normalfont B\kern-0.5em{\scshape i\kern-0.25em b}\kern-0.8em\TeX}}}

\setcopyright{none}
\makeatletter                   
\let\@authorsaddresses\@empty   
\makeatother                    %

\usepackage{listings,textcomp}
\usepackage{xcolor}

\usepackage{enumerate}
\usepackage{tabularx}

\newcommand{\val}{\ensuremath{\mathsf{Val}}}
\newcommand{\cL}{\ensuremath{\mathcal{L}}}
\newcommand{\cP}{\ensuremath{\mathcal{P}}}

\makeatletter
\lstnewenvironment{gql}[1][]{%
    \noindent\minipage[b]{\linewidth}%
    \centering%
    \tabular{@{}c@{}}%
}{\endtabular\endminipage\vspace{5pt}}
\makeatother

\usepackage{textcomp}
\newcommand{\ttpcr}{\renewcommand{\ttdefault}{pcr}\ttfamily}
\newcommand{\sqlkw}[1]{%
	\text{\normalfont\small\ttpcr\bfseries\color{black} #1}}

 \newcommand{\sqltext}[1]{%
   \text{\normalfont\small\ttpcr #1}}

\lstnewenvironment{sql}[1][]{%
  \lstset{%
    language=SQL,%
    basicstyle=\small\ttpcr,%
    keywordstyle=\bfseries\color{black},%
    morekeywords={REFERENCES,IS},%
    deletekeywords={YEAR},%
  }%
  \lstset{#1}%
}{}

\newcommand{\gqltext}[1]{\sqltext{#1}}
\newcommand{\gqlkw}[1]{\sqlkw{#1}}
\let\gqlinline\lstinline

\def\stoptoken{!}
\newif\ifpathstop
\def\vmedge#1,{\if\stoptoken#1\let\vmnode\relax\else,\textcolor{red}{#1},\fi\vmnode}
\def\vmnode#1,{\if\stoptoken#1\let\vmedge\relax\else\textcolor{blue}{#1}\fi\vmedge}%

\definecolor{darkblue}{RGB}{0,56,153}
\definecolor{darkred}{RGB}{153,0,0}
\newcommand{\pathvalue}[1]{\mathsf{path}({\texteltid{#1}})}
\newcommand{\eltidfont}{\normalfont\ttfamily\color{darkblue}}
\newcommand{\texteltid}[1]{\text{{\eltidfont #1}}}

\usepackage{tikz}
\usetikzlibrary{automata}
\usetikzlibrary{chains}
\usetikzlibrary{arrows}
\usetikzlibrary{calc}

\newcommand{\uri}[1]{\text{\scriptsize {\tt\vphantom{yI}#1}}}

\newcommand{\urie}[1]{\uri{#1\,=\,}}

\renewcommand{\arraystretch}{0.97}

\usetikzlibrary{arrows,positioning,backgrounds} 
\tikzset{
    rt/.style={
		rectangle,
		fill = white,
		draw=black, 
		text centered,
		inner sep=0.5ex
		},
    rtt/.style={ 
    	rt,
    	inner sep=0.1ex
    	},
    ert/.style={ 
     	rt,
     	dashed
     	}, 
    ertt/.style={ 
        rtt,
        dashed
        }, 
    rect/.style={ 
        rectangle,
        fill = white,
        rounded corners,
        draw=black, 
        text centered,
        inner sep=0.8ex
        },
    rectw/.style={
        rect,
        draw=white
        },
    erect/.style={ 
    	rect,
    	dashed
    	},
    erectw/.style={ 
     	rectw,
     	dashed
     	},
    arrout/.style={
           ->,
           -latex,
           },
    arrin/.style={
           <-,
           latex-,
           },
    arrb/.style={
           <->,
           >=latex,
           }
}

\newcommand{\OMIT}[1]{}

\definecolor{RoyalBlue}{RGB}{65,105,255}

\begin{document}

\title{Graph Pattern Matching in GQL and SQL/PGQ}

\author{Alin Deutsch}
\affiliation{\institution{UCSD \& TigerGraph}\country{USA}}

\author{Nadime Francis}
\affiliation{\institution{LIGM, U Eiffel, CNRS}\country{France}}

\author{Alastair Green}
\affiliation{\institution{LDBC \& Birkbeck}\country{UK}}

\author{Keith Hare}
\affiliation{\institution{JCC Consulting \& Neo4j}\country{USA}}

\author{Bei Li}
\affiliation{\institution{Google}\country{USA}}

\author{Leonid Libkin}
\affiliation{\institution{U Edinburgh \& ENS-Paris \& Neo4j}\country{UK \& France}}

\author{Tobias Lindaaker}
\affiliation{\institution{DataStax}\country{Sweden}}

\author{Victor Marsault}
\affiliation{\institution{LIGM, U Eiffel, ENPC, CNRS}\country{France}}

\author{Wim Martens}
\affiliation{\institution{University of Bayreuth}\country{Germany}}

\author{Jan Michels}
\affiliation{\institution{Oracle}\country{USA}}

\author{Filip Murlak}
\affiliation{\institution{U Warsaw}\country{Poland}}

\author{Stefan Plantikow}
\affiliation{\institution{Neo4j}\country{Germany}}

\author{Petra Selmer}
\affiliation{\institution{Neo4j}\country{UK}}

\author{Hannes Voigt}
\affiliation{\institution{Neo4j}\country{Germany}}

\author{Oskar van Rest}
\affiliation{\institution{Oracle}\country{USA}}

\author{Domagoj Vrgo\v{c}}
\affiliation{\institution{PUC Chile \& IMFD}\country{Chile}}

\author{Mingxi Wu}
\affiliation{\institution{TigerGraph}\country{USA}}

\author{Fred Zemke}
\affiliation{\institution{Oracle}\country{USA}}

\renewcommand{\shortauthors}{Deutsch et al.}

\begin{abstract}
 As graph databases become widespread, JTC1---the committee in joint charge of information technology standards for the International Organization for Standardization (ISO), and International Electrotechnical Commission (IEC)---has approved a project to create GQL, a standard property graph query language. This complements
a project to extend SQL with a new part, SQL/PGQ, which specifies how to define graph views over an SQL tabular schema, and to run read-only queries against them. 

Both projects have been assigned to the ISO/IEC JTC1 SC32 working group for Database Languages, WG3, which continues to maintain and enhance SQL as a whole. This common responsibility helps enforce a policy that the identical core of both PGQ and GQL is a graph pattern matching sub-language, here termed GPML. 

The WG3 design process is also analyzed by an academic working group, part of the Linked Data Benchmark Council (LDBC), whose task is to produce a formal semantics of these graph data languages, which complements their standard specifications.

This paper, written by members of WG3 and LDBC, presents the key elements of the GPML of SQL/PGQ and GQL in advance of the publication of these new standards. 
 \end{abstract}

\OMIT{

}


\maketitle

\section{Introduction}
Graph databases are becoming widely used \cite{cacm21}. 
Many applications
require graph-structured data for analysis: they are found in a
variety of scientific domains, such as biology and
chemistry\cite{AlphaFold}. Machine learning increasingly centres
on graph networks \cite{DLgraphs-book,GL-book}. Business
transaction records, and derived metrics and control data, are
viewed as graphs to detect fraud patterns, understand market
trends and customer behaviours, and to model data lineage and
complex access control rules. 

Graphs appeal to many kinds of
users because the graph data model is akin to conceptual models,
which reflect our intuitive view of information as the properties
of entities and their relationships. Human knowledge, when
gathered incrementally as linked data items, naturally falls into
the shape of a graph: this is visible in open-source knowledge
graphs, such as Wikidata and DBpedia, and in the internal
knowledge bases that big tech companies build in order to inform
their decision and service-oriented tasks.

Two of the most popular ways of representing graphs are the  \emph{Resource Description Framework} (RDF)~\cite{rdf}, and  \emph{property graphs}~\cite{surveyChile}. In RDF, the data is modelled as a directed edge-labelled graph. On the other hand, property graph models the data as a mixed (that is, partially directed) multigraph. Both nodes and edges can be labelled, and also attributed (that is, associated with property/value pairs). In this paper we focus on property graphs.

The property graph data model has gained adoption in many commercial database systems. Examples include  AgensGraph, Amazon Neptune, ArangoDB,  CosmosDB, DataStax Enterprise Graph, HANA Graph, IBM Db2 Graph, JanusGraph, Neo4j, Oracle Server and PGX, RedisGraph, Sparksee, Stardog, and TigerGraph. 
Unlike RDF with its query language SPARQL~\cite{sparql11}, which is a W3C standard, property graph systems possess disparate storage models and querying facilities, present either in the form of APIs like Gremlin, or, since 2010, in the form of declarative graph query languages including Cypher from Neo4j \cite{Cypher}, GSQL from TigerGraph \cite{TigerGraph}, and PGQL from Oracle \cite{PGQL}, as well as industry/academia prototypes such as G-CORE \cite{gcore}. Pre-existing property graph data languages overlap substantially in core features, but each has also introduced unique innovations. The 2015 openCypher project has led to a widening industrial use of Cypher as a language for property graphs, but did not succeed on its own in establishing a standard. 

As graph data management has expanded as a product category, the lack of a standard query language (and associated sub-languages for schema definition) for property graphs has been felt more strongly. The situation resembles the early days of the relational/tabular data model, which led to the standardization of SQL, a hugely successful  lingua franca. In 2019 the Joint Technical Committee 1 of ISO/IEC, which defines information technology standards for the International Organization for Standardization, and International Electrotechnical Commission, approved a project to create GQL, a standard property graph query language with full CRUD (create/read/update/delete) and catalog capability. GQL builds on prior graph languages, as well as a new part 16 of SQL, in development since 2017, called SQL/PGQ. PGQ (short for \emph{property graph queries}) specifies how to define graph views over an SQL tabular schema, and to run read-only queries over such views, that can be projected by an SQL SELECT statement.

GQL and SQL/PGQ share a common data model, and a common \emph{graph pattern matching language}, called \emph{GPML} throughout the paper. Both language projects have been assigned to the ISO/IEC JTC1 SC32 (Subcommittee 32) Working Group for Database Languages (WG3) which continues to be responsible for maintaining and enhancing SQL as a whole. This structure serves a policy that GPML be kept identical in GQL and SQL/PGQ. The common GPML sublanguage is the central subject of this paper. 

Graph query languages extract data from a graph that matches a graph pattern. In GPML a graph pattern is a collection of path patterns, 
which when applied to a property graph, results in a set of \emph{path bindings}, each mapping variables in the expression to graph elements (node and edges) forming a path in the graph. These \emph{variable bindings} can be used to refer to the graph elements and the values of their properties. The way in which path bindings are projected to produce query results varies between SQL/PGQ and GQL, but the set of path bindings matching a path pattern for a given graph (prior to result projection) will be the same for both languages.

The standards process for both languages is governed by WG3, whose expert members represent the national standards bodies
of China, Denmark, Finland, Germany, Japan, Korea, the Netherlands, Sweden, the UK, and the USA. 
In addition WG3 has a liaison relationship with Linked Data Benchmark Council (LDBC). LDBC is a consortium of industrial companies, research institutes, academic researchers and consultants. LDBC defines benchmark standards for graph data workloads, but also supports the work of WG3 on GQL and SQL/PGQ by adding the expertise of its members to suggest improvements, or new ideas that may ultimately be incorporated in the official standards (for example, the Property Graph Schema Working Group in LDBC). As part of this process, an academic group was created under the auspices of LDBC, called FSWG (Formal Semantics Working Group). Members of this group previously provided a complete formal semantics of Cypher features \cite{Cypher}. The group's goal was to scrutinize and comment on the design of the GPML sublanguage, and to suggest improvements.

This paper gives an accessible summary of the GPML of GQL and SQL/PGQ. It does so before these two standards are published, and before vendors (including those which are represented in the author list) have released  implementations of the features of GPML---but at an advanced stage of consensus on its scope and design. 

The paper is organized as follows. In 
Section \ref{sec:propertygraphs} we review the property graph model and in Section 
\ref{sec:gpm-today} we give a brief overview of pattern matching facilities in existing languages and survey related work. 
Section \ref{sec:gpml} describes the main functionalities of the GPML of GQL and SQL/PGQ. In Section 
\ref{sec:sr} we explain how the language ensures finiteness of outputs, as there could be infinitely many cyclic paths in a graph. 
Section 
\ref{sec:detailedexample} provides a detailed explanation of the pattern-matching algorithm by means of an example. Section \ref{sec:lf} describes future plans in terms of the Standard release, industrial implementations, and new research questions that the GPML poses.

\section{Property Graphs}\label{sec:example}\label{sec:propertygraphs}
\begin{figure*}
    \resizebox{\linewidth}{!}{
    \begin{tikzpicture}
            \tikzstyle{every state}=[draw,thick,rectangle,rounded corners,fill=white!85!black,minimum size=5mm, text=black, font=\ttfamily, inner sep=0pt]
            \tikzstyle{every node}=[font=\ttfamily]

        \node[state] (acc1) at (-4,3) {
          \begin{tabular}{l}
            \small {\bf owner:} Scott\\
            \small {\bf isBlocked:} no\\
          \end{tabular}
        };

        \node[state] (acc2) at (5,3) {
          \begin{tabular}{l}
            \small {\bf owner:} Aretha\\
            \small {\bf isBlocked:} no\\
          \end{tabular}
        };

        \node[state] (acc3) at (0,0) {
          \begin{tabular}{l}
            \small {\bf owner:} Mike\\
            \small {\bf isBlocked:} no\\
          \end{tabular}
        };

        \node[state] (acc4) at (10,0) {
          \begin{tabular}{l}
            \small {\bf owner:} Jay\\
            \small {\bf isBlocked:} yes\\
          \end{tabular}
        };

        \node[state] (acc5) at (-5,-3) {
          \begin{tabular}{l}
            \small {\bf owner:} Charles\\
            \small {\bf isBlocked:} no\\
          \end{tabular}
        };

        \node[state] (acc6) at (5,-3) {
          \begin{tabular}{l}
            \small {\bf owner:} Dave\\
            \small {\bf isBlocked:} no\\
          \end{tabular}
        };

        \node[rectangle,draw,fill=white] (acc1label) at ($(acc1)+(0.5,0.6)$) {\ttfamily \small Account};
        \node[rectangle,draw,fill=white] (acc2label) at ($(acc2)+(0.5,0.6)$) {\ttfamily \small Account};
        \node[rectangle,draw,fill=white] (acc3label) at ($(acc3)+(0.5,0.6)$) {\ttfamily \small Account};
        \node[rectangle,draw,fill=white] (acc4label) at ($(acc4)+(0.5,0.6)$) {\ttfamily \small Account};
        \node[rectangle,draw,fill=white] (acc5label) at ($(acc5)+(0.5,0.6)$) {\ttfamily \small Account};
        \node[rectangle,draw,fill=white] (acc6label) at ($(acc6)+(0.5,0.6)$) {\ttfamily \small Account};
      
        \node[text=darkred] at ($(acc1)+(.1,-.6)$) {a1};
        \node[text=darkred] at ($(acc2)+(.6,-.6)$) {a2};
        \node[text=darkred] at ($(acc3)+(,-.6)$) {a3};
        \node[text=darkred] at ($(acc4)+(.1,-.6)$) {a4};
        \node[text=darkred] at ($(acc5)+(.1,-.6)$) {a5};
        \node[text=darkred] at ($(acc6)+(.1,-.6)$) {a6};

        \node[state,dashed,fill=white!95!black] (t1) at (-1.5,2) {
            \begin{tabular}{l}
                \small {\bf date:} 1/1/2020\\
                \small {\bf amount:} 8M\\
            \end{tabular}
        };
        \node[state,dashed,fill=white!95!black] (t2) at (1.5,2) {
            \begin{tabular}{l}
                \small {\bf date:} 2/1/2020\\
                \small {\bf amount:} 10M\\
            \end{tabular}
        };
        \node[state,dashed,fill=white!95!black] (t3) at (9,2) {
            \begin{tabular}{l}
                \small {\bf date:} 3/1/2020\\
                \small {\bf amount:} 10M\\
            \end{tabular}
        };
        \node[state,dashed,fill=white!95!black] (t4) at (9,-2) {
            \begin{tabular}{l}
                \small {\bf date:} 4/1/2020\\
                \small {\bf amount:} 10M\\
            \end{tabular}
        };
        \node[state,dashed,fill=white!95!black] (t5) at (1.7,-2) {
            \begin{tabular}{l}
                \small {\bf date:} 6/1/2020\\
                \small {\bf amount:} 10M\\
            \end{tabular}
        };
        \node[state,dashed,fill=white!95!black] (t6) at (0,-3.5) {
            \begin{tabular}{l}
                \small {\bf date:} 7/1/2020\\
                \small {\bf amount:} 4M\\
            \end{tabular}
        };
        \node[state,dashed,fill=white!95!black] (t7) at (-1.2,-2) {
            \begin{tabular}{l}
                \small {\bf date:} 8/1/2020\\
                \small {\bf amount:} 6M\\
            \end{tabular}
        };
        \node[state,dashed,fill=white!95!black] (t8) at (-5,0) {
            \begin{tabular}{l}
                \small {\bf date:} 9/1/2020\\
                \small {\bf amount:} 9M\\
            \end{tabular}
        };
      
        \node[rectangle,draw,dashed,fill=white] (t1label) at ($(t1)+(0.5,0.6)$) {\ttfamily \small Transfer};
        \node[rectangle,draw,dashed,fill=white] (t2label) at ($(t2)+(0.5,0.6)$) {\ttfamily \small Transfer};
        \node[rectangle,draw,dashed,fill=white] (t3label) at ($(t3)+(0.5,0.6)$) {\ttfamily \small Transfer};
        \node[rectangle,draw,dashed,fill=white] (t4label) at ($(t4)+(0.5,0.6)$) {\ttfamily \small Transfer};
        \node[rectangle,draw,dashed,fill=white] (t5label) at ($(t5)+(0.5,0.6)$) {\ttfamily \small Transfer};
        \node[rectangle,draw,dashed,fill=white] (t6label) at ($(t6)+(0.5,0.6)$) {\ttfamily \small Transfer};
        \node[rectangle,draw,dashed,fill=white] (t7label) at ($(t7)+(0.5,0.6)$) {\ttfamily \small Transfer};
        \node[rectangle,draw,dashed,fill=white] (t8label) at ($(t8)+(0.5,0.6)$) {\ttfamily \small Transfer};

        \node[text=darkblue] at ($(t1)+(.1,-.6)$) {t1};
        \node[text=darkblue] at ($(t2)+(.1,-.6)$) {t2};
        \node[text=darkblue] at ($(t3)+(-.1,-.6)$) {t3};
        \node[text=darkblue] at ($(t4)+(-1.5,0)$) {t4};
        \node[text=darkblue] at ($(t5)+(.1,-.6)$) {t5};
        \node[text=darkblue] at ($(t6)+(1.45,.2)$) {t6};
        \node[text=darkblue] at ($(t7)+(.1,-.6)$) {t7};
        \node[text=darkblue] at ($(t8)+(.1,-.6)$) {t8};

        \node[state] (p1) at (-9,-1) {
          \begin{tabular}{l}
            \small {\bf number:} 111\\
            \small {\bf isBlocked:} no\\
          \end{tabular}
        };

        \node[state] (p2) at (3.5,1) {
          \begin{tabular}{l}
            \small {\bf number:} 222\\
            \small {\bf isBlocked:} no\\
          \end{tabular}
        };

        \node[state] (p3) at (13,0) {
          \begin{tabular}{l}
            \small {\bf number:} 333\\
            \small {\bf isBlocked:} no\\
          \end{tabular}
        };

        \node[state] (p4) at (5,-1) {
          \begin{tabular}{l}
            \small {\bf number:} 444\\
            \small {\bf isBlocked:} no\\
          \end{tabular}
        };

      \node[rectangle,draw,fill=white] (p1label) at ($(p1)+(0.45,0.6)$) {\ttfamily \small Phone};
      \node[rectangle,draw,fill=white] (p2label) at ($(p2)+(0.45,0.6)$) {\ttfamily \small Phone};
      \node[rectangle,draw,fill=white] (p3label) at ($(p3)+(0.45,0.6)$) {\ttfamily \small Phone};
      \node[rectangle,draw,fill=white] (p4label) at ($(p4)+(0.45,0.6)$) {\ttfamily \small Phone};

        \node[text=darkred] at ($(p1)+(.1,-.6)$) {p1};
        \node[text=darkred] at ($(p2)+(.1,-.6)$) {p2};
        \node[text=darkred] at ($(p3)+(.1,-.6)$) {p3};
        \node[text=darkred] at ($(p4)+(.3,-.6)$) {p4};

        \node[state] (ip1) at (-9,1) {
          \begin{tabular}{l}
            \small {\bf number:} 123.111\\
            \small {\bf isBlocked:} no\\
          \end{tabular}
        };

        \node[state] (ip2) at (-9,-3) {
          \begin{tabular}{l}
            \small {\bf number:} 123.222\\
            \small {\bf isBlocked:} no\\
          \end{tabular}
        };

      \node[rectangle,draw,fill=white] (ip1label) at ($(ip1)+(0.45,0.6)$) {\ttfamily \small IP};
      \node[rectangle,draw,fill=white] (ip2label) at ($(ip2)+(0.45,0.6)$) {\ttfamily \small IP};
      
        \node[text=darkred] at ($(ip1)+(.1,-.6)$) {ip1};
        \node[text=darkred] at ($(ip2)+(.1,-.6)$) {ip2};

        \node[state] (c1) at (-8,3) {
          \begin{tabular}{l}
            \small {\bf name:} Zembla\\
          \end{tabular}
        };
      
        \node[state] (c2) at (12.5,3) {
          \begin{tabular}{l}
            \small {\bf name:} Ankh-Morpork\\
          \end{tabular}
        };

        \node[rectangle,draw,fill=white] (c1label) at ($(c1)+(0.5,0.4)$) {\ttfamily \small Country};
        \node[rectangle,draw,fill=white] (c2label) at ($(c2)+(0.5,0.4)$) {\ttfamily \small City, Country};

        \node[text=darkred] at ($(c1)+(-.1,-.5)$) {c1};
        \node[text=darkred] at ($(c2)+(.3,-.5)$) {c2};

    \begin{pgfonlayer}{background}        
        \path[-latex,thick]
        (acc1) edge[bend left] (acc3)
        (acc3) edge[bend left] (acc2)
        (acc2) edge[bend left] (acc4)
        (acc4) edge[bend left] (acc6)
        (acc6) edge[out=190,in=-10] (acc5)
        (acc6) edge[bend left] (acc3)
        (acc3) edge[bend left] (acc5)
        (acc5) edge[bend left] (acc1)
        ;
          
        \path[-,thick,densely dotted]
        (acc1) edge node [above,sloped,text=darkblue] {hp1} (p1)
        (acc2) edge node [right,text=darkblue] {hp2} (p2)
        (acc3) edge node [above,text=darkblue] {hp3} (p2)
        (acc4) edge node [below,text=darkblue] {hp4} (p3)
        (acc5) edge node [above,sloped,text=darkblue] {hp5} (p1)
        (acc6) edge node [left,text=darkblue] {hp6} (p4)
        ;
        
        \path[-latex,thick,dotted] 
        (acc1) edge [out=180,in=0] node [above,sloped,text=darkblue] {li1} (c1)
        (acc2) edge [out=10,in=170] node [above,sloped,text=darkblue] {li2} (c2)
        (acc3) edge [out=160,in=-30] node [above,sloped,near start,text=darkblue] {li3} (c1)
        (acc4) edge [out=20,in=270] node [above,sloped,text=darkblue] {li4} (c2)
        (acc5) edge [out=140,in=290] node [above,sloped,text=darkblue] {li5} (c1)
        (acc6) edge [out=60,in=220] node [above,sloped,text=darkblue] {li6} (c2)
        ;
        
        \path[-,thick,dashed]
        (acc1) edge node [above,sloped,near start,text=darkblue] {sip1} (ip1)
        (acc5) edge node [below,text=darkblue] {sip2} (ip2)
        ;
    \end{pgfonlayer}

    \node at (12,-1.8) {Edge labels:};
    
    \path[-latex,thick] (11,-2.3) edge (12,-2.3);
    \node at (12.8,-2.3) {\small{\texttt{Transfer}}};

    \path[-latex,thick, dotted] (11,-2.7) edge (12,-2.7);
    \node at (13,-2.7) {\small{\texttt{isLocatedIn}}};

    \path[-,thick,densely dotted] (11,-3.1) edge (12,-3.1);
    \node at (12.8,-3.1) {\small{\texttt{hasPhone}}};

    \path[-,thick,dashed] (11,-3.5) edge (12,-3.5);
    \node at (13.1,-3.5) {\small{\texttt{signInWithIP}}};
    \end{tikzpicture}
    }
    \caption{A property graph with information on bank accounts, their location, and financial transations.\label{fig:propertygraph}}
\end{figure*}
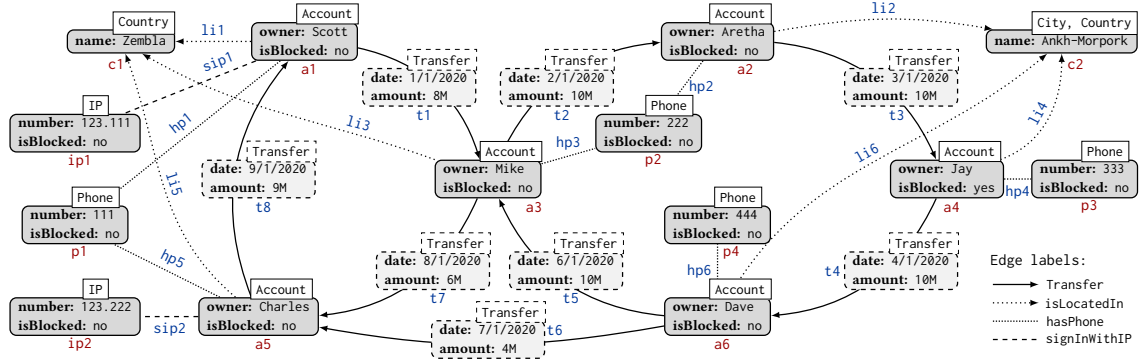

To describe property graphs, we use a common application scenario involving banking and financial transactions (used, e.g., for fraud detection). Figure~\ref{fig:propertygraph} is an example of a property graph containing information about bank accounts, their location, their associated phone numbers and IP addresses, and financial transactions between them. The graph contains \emph{nodes}, which are connected by \emph{edges}. Nodes and edges are identified by \emph{node identifiers} (\texteltid{a1}, \dots, \texteltid{a6}, \texteltid{c1}, \texteltid{c2}, \texteltid{p1},\dots,\texteltid{p4}, \texteltid{ip1}, \texteltid{ip2}) and \emph{edge identifiers} (\texteltid{t1},\dots,\texteltid{t8}, \texteltid{li1},\dots,\texteltid{li6},\texteltid{hp1},\dots,\texteltid{hp6},\texteltid{sip1},\texteltid{sip2}), respectively. Furthermore, both nodes and edges can carry \emph{labels} (e.g., Account, City, Transfer, isLocatedIn) and \emph{property/value pairs} (e.g., owner/Mike, isBlocked/no). In figures, we will depict node information in solid boxes, and edge information in dashed boxes. We use the umbrella term \emph{element} to refer to nodes or edges. When clear from the context,  we do not distinguish between node identifiers and nodes (similarly for edges).

In graph-theoretic literature, graphs are usually defined as pairs $\langle V,E\rangle$ of vertices $V$ and edges $E$ which are either two-element subsets of $V$ for undirected graphs, or pairs of vertices for directed graphs. In contrast, property graphs are \emph{multigraphs} (there can be multiple edges between two endpoint nodes);  \emph{pseudographs} (there can be an edge looping from a node to itself); they are {\em mixed}, or \emph{partially directed}: an edge can be \emph{undirected}, or can have source and target nodes, in which case it is \emph{directed} from the source to the target. They are also \emph{attributed}: graph elements can have attributes (a disjoint union of labels and properties).

This is summarized in the following definition. 
Assume that $\cL$, $\cP$, and $\val$ are countably infinite sets, containing \emph{labels},  \emph{property names}, and \emph{property values}, respectively.
\begin{definition}[Property Graph]\label{def:propertygraph}
A \emph{property graph} is defined as a tuple $G = (N, E, \rho, \lambda, \pi)$ where:
\begin{itemize}
\item $N$ is a finite set of node identifiers;
\item $E$ is a finite set of edge identifiers;
such that $N \cap E = \emptyset$;
\item $\rho : E \rightarrow (N \times N) \cup \{\{u,v\} \mid u,v \in N\}$ is a total function mapping edges 
to ordered or unordered pairs of nodes; 
\item $\lambda : (N \cup E) \rightarrow 2^{\cL}$ is a total function mapping node and edge identifiers 
to sets of labels (including the empty set);
\item $\pi : (N \cup E) \times \cP \rightharpoonup \val$ is a partial function mapping elements and property names to property values.
\end{itemize}
\end{definition}
We call an edge $e$ \emph{directed} if $\rho(e) \in N \times N$ and \emph{undirected} if $\rho(e) \in \{\{u,v\} \mid u,v \in N\}$. In both cases we say that $e$ connects $u$ and $v$. Notice that both directed and undirected edges allow $u = v$, in which case the edge is a \emph{self-loop}. Furthermore, the definition allows both directed and undirected edges to carry labels and data. Indeed, $\lambda$ and $\pi$ can be defined for directed and  undirected edges. The definition does not preclude having two different edges connecting the same nodes. 
For more details we refer the reader to~\cite{LDBC:TR:TR-2021-01}.

A property graph has a \emph{graph representation}, which is illustrated in Figure~\ref{fig:propertygraph}, but it also has a \emph{tabular representation}, which is illustrated in Figure~\ref{fig:propertygraphtable}. In the graph representation, we depicted the nodes in red and the edges in blue. We depicted the properties and values associated to nodes in grey rounded rectangles (which are dashed for edges) and most of the labels in white rectangles. To avoid clutter however, we have omitted the labels on some edges (namely \gqltext{isLocatedIn}, \gqltext{hasPhone}, and \gqltext{signInWithIP}).
The tabular representation has a relation for every combination of labels that appears on some node or edge in the graph. For instance, every label that appears in Figure~\ref{fig:propertygraph} is a relation name in the tabular representation, except \gqltext{City}, which does not appear by itself. It does appear together with Country (on node \texteltid{c2}), so the tabular representation has a relation named CityCountry, which contains node \texteltid{c2}.

\begin{figure*}[tb]
\begin{tabular}[t]{lll}
\multicolumn{3}{l}{Account}\\
\toprule
ID & owner & isBlocked\\
\midrule
\texteltid{a1} & Scott & no\\
\texteltid{a2} & Aretha & no\\
\texteltid{a3} & Mike & no\\
\dots\\
  \bottomrule
\end{tabular}
\qquad
\begin{tabular}[t]{lllll}
    \multicolumn{5}{l}{Transfer}\\
    \toprule
     ID & A\_ID1 & A\_ID2 & date & amount\\
     \midrule
     \texteltid{t1} & \texteltid{a1} & \texteltid{a3} & 1/1/2020 & 8M\\
     \texteltid{t2} & \texteltid{a3} & \texteltid{a2} & 2/1/2020 & 10M\\
     \texteltid{t3} & \texteltid{a2} & \texteltid{a4} & 3/1/2020 & 10M\\
     \dots\\
  \bottomrule
\end{tabular}

\bigskip \bigskip
\begin{tabular}[t]{lll}
  \multicolumn{3}{l}{signInWithIP}\\
  \toprule
  ID & A\_ID & s\_ID\\
  \midrule
  \texteltid{sip1} & \texteltid{a1} & \texteltid{ip1}\\
  \texteltid{sip2} & \texteltid{a5} & \texteltid{ip2}\\
  \bottomrule
\end{tabular}
\qquad
\begin{tabular}[t]{ll}
    \multicolumn{2}{l}{Country}\\
    \toprule
    ID & name\\
    \midrule
    \texteltid{c1} & Zembla\\
  \bottomrule
\end{tabular}
\qquad
\begin{tabular}[t]{ll}
    \multicolumn{2}{l}{CityCountry}\\
    \toprule
    ID & name\\
    \midrule
    \texteltid{c2} & Ankh-Morpork\\
  \bottomrule
\end{tabular}
\caption{Excerpts of five tables of the tabular representation of the property graph in Figure~\ref{fig:propertygraph}. The tables Transfer and signInWithIP represent edges, the others represent nodes.\label{fig:propertygraphtable}}
\end{figure*}

A \emph{path} is an alternating sequence of nodes and edges such that: (1)~it starts and ends with a node; and (2)~subsequent nodes in the sequence are connected by the edge between them.\footnote{As is usual in the graph database literature \cite{openCypher,MendelzonW95,Woo,B13}, we use the term path to denote what is called \emph{walk} in the graph theory literature \cite{bollobas2013modern}.} 
We write paths as $\pathvalue{c1,li1,a1,t1,a3,hp3,p2}$
indicating a path from node \texteltid{c1} to node \texteltid{p2} that first follows edge \texteltid{li1} in reverse direction, then the edge \texteltid{t1} in forward direction, and then the undirected edge \texteltid{hp3}.





\section{Graph Pattern Matching Today}\label{sec:gpm-today}

\label{sec:existing}
As identified in a recent survey of modern graph query languages~\cite{surveyChile}, when querying graph databases, one usually starts by selecting atomic graph elements, such as nodes, edges, or paths. Coming back to our example from Figure \ref{fig:propertygraph}, basic elements we might want to select are, for instance:
\begin{itemize}
\item All the nodes representing a blocked account;
\item All the edges representing a money transfer from a blocked to a non-blocked account that occurred on a specific date; 
\item All the paths whose edges represent money transfers, and that start in a non-blocked account, but end in a blocked one.
\end{itemize}

\begin{figure} 
\begin{center}
   \begin{tikzpicture}
        \tikzstyle{every state}=[draw,thick,rectangle,rounded corners,fill=white!85!black,minimum size=5mm, text=black, font=\ttfamily, inner sep=0pt]

        \node[state] (n1) at (0,0) {
          \begin{tabular}{l}
            \urie{isBlocked}\uri{yes}\\
          \end{tabular}
        };
        \node[rt] (n1label) at ($(n1)+(0,0.35)$) {\uri{$x$ : Account}};
        
        \node[state] (n2) at (-3,-1.3) {
          \begin{tabular}{l}
            \urie{isBlocked}\uri{yes}\\
          \end{tabular}
        };
        \node[rt] (n2label) at ($(n2)+(0,0.35)$) {\uri{$x$ : Account}};
        
		\node[state] (n3) at (3,-1.3) {
          \begin{tabular}{l}
            \urie{isBlocked}\uri{no}\\
          \end{tabular}
		};
        \node[rt] (n3label) at ($(n3)+(0,0.35)$) {\uri{$y$ : Account}};

		\draw[arrout] (n2) to 
		node[state,fill=white!95!black,dashed] (e1) {
          \begin{tabular}{l}
            \urie{date}\uri{3/1/2020}\\
          \end{tabular}
    	} 
		(n3);
        \node[ert] (e1label) at ($(e1)+(0,0.35)$) {\uri{$e$ : Transfer}};

		\node[state] (n4) at (-3,-2.6) {
          \begin{tabular}{l}
            \urie{isBlocked}\uri{no}\\
          \end{tabular}
        };
        \node[rt] (n4label) at ($(n4)+(0,0.35)$) {\uri{$x$ : Account}};

		\node[state] (n5) at (3,-2.6) {
          \begin{tabular}{l}
            \urie{isBlocked}\uri{yes}\\
          \end{tabular}
        };
        \node[rt] (n5label) at ($(n5)+(0,0.35)$) {\uri{$y$ : Account}};
        
		\draw[arrout] (n4) to 
		node[erect, below] (e2) {
			} 
		(n5);
		
		\node[ert] (le2) at (e2.north) {\uri{:Transfer}$^+$};	
		
		\node at (0,-.5) {{\small (a)}};
		\node at (0,-1.8) {{\small (b)}};
		\node at (0,-3.05) {{\small (c)}};
    \end{tikzpicture}
\end{center}
\caption{Node pattern, edge pattern, and a path pattern.}
\label{fig:atoms} 		
\end{figure}
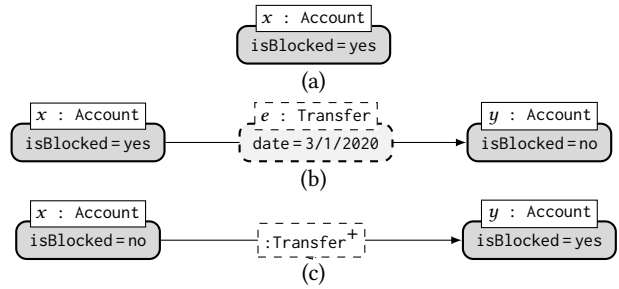

In an abstract sense, we could represent these elements as patterns shown in Figure \ref{fig:atoms}. 
In addition to standard elements of a property graph, these graph patterns also attach variables $x$, $y$, $e$, \mbox{etc.} to different parts of the patterns. To match such a pattern in a property graph, we need a mapping that links the node patterns to nodes in the graph, and similarly with edge and path patterns. For instance, the pattern (a) from Figure \ref{fig:atoms} matches to an \texttt{Account} node such that the value of the property \texttt{isBlocked} is set to \texttt{yes}. Similarly, the pattern (b) of Figure \ref{fig:atoms} matches the node variables $x$ and $y$ to two accounts (the first one blocked, and the second one not), while $e$ matches an edge of type \texttt{Transfer} connecting these two nodes, with the value of \texttt{date} set to \texttt{3/1/2020}. 

The most interesting pattern, shown in Figure \ref{fig:atoms}(c), specifies a path of arbitrary length. Starting from \cite{MendelzonW89}, specifying paths via regular expressions over edge types has been well established both in the research community \cite{B13,Woo,CGLV03,surveyChile} and in the industry \cite{PS08,Cypher, PGQL,TigerGraph}. Several ways of defining valid paths (simple, arbitrary, etc.) have been proposed \cite{MendelzonW95,Cypher}; we discuss them in later sections. 

\begin{figure} 
\begin{center}
	\begin{tikzpicture}
        \tikzstyle{every state}=[draw,thick,rectangle,rounded corners,fill=white!90!black,minimum size=5mm, text=black, font=\ttfamily, inner sep=0pt]
        \tikzstyle{every node}=[font=\ttfamily]
		\node[state] (n1) {
          \begin{tabular}{l}
            \urie{name}\uri{Ankh-Morpork}\\
          \end{tabular}
        };
        \node[rt] (n1label) at ($(n1)+(0,0.35)$) {\uri{:City}};
		
		\node[state] (n2) at (-3,-1.5) {
          \begin{tabular}{l}
            \urie{isBlocked}\uri{no}\\
          \end{tabular}
        };

		\node[rt] (n2label) at ($(n2)+(0,0.35)$) {\uri{$x$ $:$ Account}};
		
		\draw[arrout] (n2label) to 
		node[below, erectw] (e1) {} 
		(n1);
		
		\node[ert] (le1) at (e1.north) {\uri{:isLocatedIn}};

		\node[state] (n5) at (3,-1.5) {
          \begin{tabular}{l}
            \urie{isBlocked}\uri{yes}\\
          \end{tabular}
        };

		\node[rt] (n5label) at ($(n5)+(0,0.35)$) {\uri{$y$ $:$ Account}};

		\draw[arrin] (n1) to
		node[ert] (e6) {\uri{:isLocatedIn}
			}
		(n5label);		

		\draw[arrin] (n5) to
		node[ert] (e) {\uri{:Transfer}$^+$
			}
		(n2);		

	\end{tikzpicture}
\end{center}
\caption{Pattern of fraudulent accounts in Ankh-Morpork.}
\label{fig:pattern}	
\end{figure}
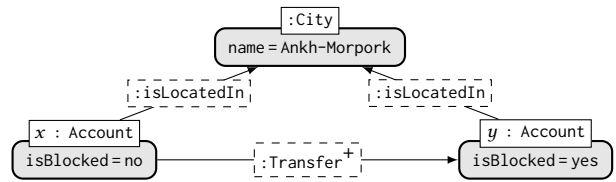

With these basic building blocks at hand, we can define more complex graph patterns. For instance, assume that in the property graph from Figure \ref{fig:propertygraph}, we want to identify all pairs of owners with accounts that are located in Ankh-Morpork, and are connected by a path of (arbitrarily many) money transfers. Additionally, we might want to stipulate that the first account is blocked, while the second one is not blocked. Following our graphical representation, one can visualize this query as the graph pattern in Figure \ref{fig:pattern}. Notice that the pattern only specifies what we want to match in the graph, not what we want to output (the owners of \gqltext{x} and \gqltext{y}).

In general, a graph pattern specifies a set of nodes and their connections via edges and paths, filtered according to labels and values of some of their properties. Such a graph-shaped query gets matched to the graph database  to obtain a (collection of) result(s).
Graph patterns form the core of most existing graph query languages, and have been present in academic research (under the name of conjunctive regular path queries, or CRPQs \cite{CGLVkr,B13,Woo}), in W3C recommendations (\mbox{e.g.} SPARQL \cite{PS08}), and in existing property graph engines \cite{Cypher,openCypher,TigerGraph,PGQL}.
We next summarize the support for graph patterns in SPARQL, Neo4j's Cypher, Oracle PGQL, and TigerGraph's GSQL. For historical reasons, we begin with SPARQL.

The SPARQL standard \cite{PS08} prescribes graph pattern matching on edge-labeled graphs represented as a single ternary relation. Neither labels nor properties are present in SPARQL, and  matching is done only between nodes and constants. Path queries are supported by means of property paths \cite{sparql11} that permit full 2RPQs \cite{CGLV03} under arbitrary path semantics. In order to avoid infinite results when cycles are present, SPARQL prescribes that one can only check for the existence of a path between a pair of nodes, but cannot count the number of such paths, or try to reconstruct them \cite{ACP12,LM13}. A much simplified version of the query from Figure \ref{fig:pattern} in SPARQL (assuming cities are identified by their names), would be the following:

\begin{gql}
SELECT ?x, ?y
WHERE { ?x isLocatedIn "Ankh-Morpork".
        ?y isLocatedIn "Ankh-Morpork".
        ?x Transfer+ ?y }
\end{gql}

The most common query language for property graphs is Cypher \cite{Cypher} that originated at Neo4j and has since been implemented by vendors such as Amazon, Agens Graph, Katana Graph, Memgraph, RedisGraph, and SAP HANA \cite{openCypherProjects}.
The owners of the accounts in the query from Figure \ref{fig:pattern} can be found with Cypher as follows: 

\begin{gql}
MATCH (a:Account {isBlocked:'no'})-[:isLocatedIn]->
      (g:City {name:'Ankh-Morpork'})<-[:isLocatedIn]-
      (b:Account {isBlocked:'yes'}), 
  p = (a)-[:Transfer*1..]->(b)
RETURN a.owner, b.owner
\end{gql}

\noindent The query first matches the variables \gqltext{a} and \gqltext{b} to an  unblocked and a blocked account located in Ankh-Morpork, using its ``ASCII-art'' syntax for specifying edges. It then matches \gqltext{p} to a whole path of transfers of arbitrary length from \gqltext{a} to \gqltext{b}, and returns the owners.
Unlike SPARQL, Cypher allows returning  paths: we can return \gqltext{p} simply by including it in the \gqlkw{RETURN} clause.

Cypher's pattern matching has multiple other features. 
It permits label disjunctions  and inline predicates in variable-length path patterns: \gqlkw{MATCH} \gqlinline!(a)-[:X|Y*{weight:1}]->(b)!. It supports testing for the presence or absence of a path relative to an element specified in a match: 
\gqlkw{MATCH} \gqlinline!(a:Person)->(:Cat)! \gqlkw{WHERE NOT} \gqlinline!(a)->(:Dog)!. 
It also allows custom graph traversals by calling user-defined procedures via \gqlkw{CALL}, and complex predicates involving nested data, especially paths.
Cypher also allows additional operations on matched paths such as returning a single shortest or all shortest paths, returning all nodes or edges in a path,
the length of a path, as well as the number of paths found. 

PGQL (Property Graph Query Language)~\cite{van2016pgql} is an SQL-like graph pattern matching query language that is open-sourced by Oracle.
Where features overlap, PGQL follows SQL's syntax and semantics, such as in \gqlkw{SELECT}, \gqlkw{WHERE}, \gqlkw{GROUP} \gqlkw{BY}, \gqlkw{HAVING}, and \gqlkw{ORDER BY}, as well as for functions, aggregations, predicates, existential and scalar subqueries.
The core of a PGQL query is the graph pattern, which is spread over multiple \gqlkw{MATCH} clauses that are syntactically placed inside the \gqlkw{FROM} clause.
The owners of the accounts of query in Figure \ref{fig:pattern} can be found as follows.

\begin{gql}
SELECT x.owner AS A, y.owner AS B
FROM
  MATCH (x:Account)-[:isLocatedIn]->
        (g:City)<-[:isLocatedIn]-(y:Account),
  MATCH ANY (x)-[e:Transfer]->+(y)
WHERE x.isBlocked='no' AND y.isBlocked='yes'
  AND g.name='Ankh-Morpork'
\end{gql}

\noindent We can return the whole path of \gqltext{Transfer} edges between \gqltext{x} and \gqltext{y} by replacing the first line of the query with \gqlkw{SELECT} \gqltext{x.owner} \gqlkw{AS} \gqltext{A,} \gqltext{y.owner} \gqlkw{AS} \gqltext{B,} \gqltext{LISTAGG(e.ID,\:', ')}, assuming that \gqltext{e.ID} is the ID of edge \gqltext{e}. 
Here, the variable \gqltext{e} is treated as a \emph{group variable} that matches subsequent edges of the path and can be used to aggregate data along paths of variable length. For example, one can compute the length of the path using \gqltext{COUNT(e)}, or filter out paths with repeated edges using \gqlkw{WHERE}~\gqltext{COUNT(e)\:=\:COUNT(DISTINCT\:e)}. The aggregate \gqltext{LISTAGG(e.ID,\:', ')} produces a comma-separated list of values encoded as a single string of characters.
\gqlkw{ANY} is used to obtain an arbitrary single
path from \gqltext{x} to \gqltext{y}. 
PGQL supports also 
\gqlkw{ANY\:SHORTEST}, 
\gqlkw{ALL\:SHORTEST},
\gqlkw{TOP}\:\gqltext{k}\:\gqlkw{SHORTEST}, 
\gqlkw{ANY\:CHEAPEST}, and
\gqlkw{TOP} \gqltext{k} \gqlkw{CHEAPEST},
allowing for retrieving multiple paths between a pair of nodes. \gqlkw{ALL} gives all paths but requires an upper bound on the path length, \mbox{e.g.}, \gqlinline!{1,4}! instead of \gqltext{+}.

\sloppy
The design philosophy behind 
TigerGraph's GSQL (Graph SQL) language~\cite{TigerGraph,tigergraphTR} is to flatten the learning curve for the largest community of potential adopters, namely SQL programmers.
Like PGQL, GSQL supports
SQL-style \gqlkw{SELECT}, \gqlkw{WHERE}, 
\gqlkw{GROUP} \gqlkw{BY}, \gqlkw{HAVING}, \gqlkw{LIMIT} and \gqlkw{ORDER BY} clauses, aggregation, and graph
patterns in the \gqlkw{FROM} clause.
The running example query from 
Figure~\ref{fig:pattern} is expressed in GSQL as follows, using table
\gqltext{T} to hold the result.

\begin{gql}
SELECT x.owner AS A, y.owner AS B INTO T
FROM Account:x -(isLocatedIn>)- City:g
     -(<isLocatedIn)- Account:y,
     :x -(Transfer>+)- :y
WHERE x.isBlocked='no' AND y.isBlocked='yes'
  AND g.name='Ankh-Morpork'
GROUP BY A, B
\end{gql}

\noindent
GSQL's default semantics is \gqlkw{ALL\:SHORTEST},
hence there is no upper bound on the \gqltext{+} quantifier.
While supporting group and path variables is on the near-future roadmap for GSQL, this is merely a matter of adding syntactic sugar as they 
are currently expressible using a GSQL-specific aggregation paradigm
based on {\em accumulators}. These are containers that 
can be attached at query time to the vertices and written to during the pattern matching phase. 
Accumulator inputs are aggregated via pre- or user-defined 
binary operators.
Accumulator-based aggregation was shown 
to enjoy expressivity and efficiency benefits over SQL-style aggregation \cite{DeutschXWL20}.


\section{Graph pattern matching language}\label{sec:gpml}
Recall that we use the acronym GPML to denote 
the common graph pattern matching language 
that is shared by SQL/PGQ and GQL. In the database domain there is no
standard or product by those initials to our knowledge;
it is just a convenient abbreviation for this paper. This section is devoted to explaining how GPML works. In general, a GPML pattern is an expression of the form:

\begin{gql}
MATCH °{\normalfont\textit{Pattern}}°
\end{gql}

\noindent where \textit{Pattern} is a pattern specifying the parts of our input property graph we want to explore. The \gqlkw{MATCH} clause may be optionally followed by a \gqlkw{WHERE} clause to filter the results; we discuss it later in the section. For example, the following expression 
retrieves all the accounts in our graph from Figure~\ref{fig:propertygraph} that are not blocked:

\begin{gql}
MATCH (x:Account WHERE x.isBlocked='no')
\end{gql}

\noindent Intuitively, this pattern, which is an example of a \emph{node pattern}, binds to \gqltext{x} all the nodes in the graph which have the label \gqltext{Account}, and whose \gqltext{isBlocked} property has the value \gqltext{no}. Another basic pattern, called the \emph{edge pattern}  extracts edges from our graph. For instance, the following pattern asks for all the edges which represent transfers with a value of more than 5 million:

\begin{gql}
MATCH -[e:Transfer WHERE e.amount>5M]->
\end{gql}

\noindent We shall use shorthands such as 5M for readability. The variable \gqltext{e} gets bound to an edge in our graph, giving us access to its data. We later explain how one can also retrieve the endpoints of the edge. Generally, node and edge patterns retrieve graph elements -- nodes and edges -- and can then be combined in more complex ways to define paths and patterns present in our graph. In what follows we specify how all of this is supported in GPML. We start by expanding on our definition of node and edge patterns.

\subsection{Accessing Nodes and Edges}

\paragraph*{Node patterns.} The most basic query pattern allows the user to fetch nodes inside a property graph. For instance, to extract all of the nodes of the graph, the user can write the following\footnote{The ascii art \gqlinline!()! used for nodes is meant to be suggestive of how nodes are usually drawn as circles in visual representation of a graph.}

\begin{gql}
MATCH (x)
\end{gql}

The variable \gqltext{x} here is called an \emph{element variable}, since it binds to a graph element. In the case of~\gqltext{x}, it will be bound to a node in the graph, since it is placed inside the \gqlinline!( )! braces, which signify that we are talking about a node. 
For instance, if evaluated on the property graph from Section \ref{sec:example}, this query will return bindings that map \gqltext{x} to accounts, cities, phones, and IPs.\footnote{For now we model query evaluation as a process taking a property graph and a query pattern as input, and producing as output a multiset of  \emph{bindings} from the set of variables to the set containing graph elements and property values. In Section \ref{sec:detailedexample} we extend this to capture more advanced objects such as arbitrary length paths.}

Of course, one rarely wants to obtain all the nodes in a graph, and usually will want to restrict the obtained results in some way. A natural way to do this is to look only for nodes with a specific label. GPML does this via \emph{label expressions}. For instance, 
to return nodes that correspond to accounts, one writes:

\begin{gql}
MATCH (x:Account)
\end{gql}

\noindent Here the label is explicitly specified, and only nodes with the label \gqltext{Account} will now be bound to~\gqltext{x}. More generally, label expressions allow conjunctions (\gqltext{\&}), disjunctions (\gqltext{|}), negations (\gqltext{!}), and grouping of individual labels by using parentheses. For instance, to capture all the nodes that are either accounts, or IP addresses, we would write \gqlinline{MATCH} \gqlinline{(x:Account|IP)}. There is also a wildcard symbol \gqltext{\%} matching any label; e.g., pattern 
\gqlinline|(:!
matches nodes that have no labels. 

Further filters on a node restrict the values of some of its properties. As seen earlier, if we are interested only in those accounts that are not blocked, the pattern \gqlinline!(x:Account)! changes to 
\gqlinline!(x:Account WHERE x.isBlocked='no')!.
The \gqlkw{WHERE} clause can be put outside \gqlkw{MATCH} to be used as a postfilter for produced matches:

\begin{gql}
MATCH (x:Account)
WHERE x.isBlocked='no'
\end{gql}

\noindent The \gqlkw{WHERE} clause can support a host of search conditions, and these may be combined into logical statements using \gqlkw{AND}, \gqlkw{OR}, and \gqlkw{NOT}.

We note that each of the restricting elements in a node pattern (e.g. the variable, the label specification, or the \gqlkw{WHERE} condition) is optional. That is, any of them can be either present or absent. This makes the following the simplest possible node pattern: \gqlinline!MATCH ()!. 
Since there is no variable, there is no syntax 
to reference the nodes bound to this pattern; consequently
it is of little use standing alone as the sole element
pattern as in this example. However, below we explain how such a pattern can be combined with edge patterns in order to extract paths and patterns in the graph. Intuitively, this construct will allow us to have a placeholder for any node in the graph, thus allowing us to link it with other graph elements.
It is of more use when
concatenated with edge patterns, as in 

\begin{gql}
MATCH (x)-[:Transfer]->()-[:isLocatedIn]->(y)
\end{gql}

In the preceding example, the user is only interested 
in the bindings to \gqltext{x} and \gqltext{y}, consequently the user 
omitted the variables from the two edge patterns 
and the middle node pattern.

\paragraph*{Edge patterns.} These let us explore an edge connecting two nodes. A basic way of doing this is via the pattern\footnote{Here again we use the  ascii art \gqlinline!-[]->! to mimic how one draws an edge in a graph.} 

\begin{gql}
MATCH -[e]->
\end{gql}

\noindent which searches all the directed edges in the graph, 
and binds them to the element variable \gqltext{e}. That is, the bindings generated by this pattern will map \gqltext{e} to any identifier of a directed edge in the graph.  
If we wish to use specific types of edges (\mbox{e.g.} undirected), this can be done by special symbols used in the edge pattern. For example, all undirected edges 
can be recovered by \lstinline{MATCH ~[e]~}.

The full specification of possible edge direction restrictions and their combinations is provided in Figure~\ref{fig.edgepatterns}. In the figure, \emph{spec} is an optional element of the form \gqltext{e:labelExpr} \gqlkw{WHERE} \gqltext{condition}, where \gqltext{labelExpr} and \gqltext{condition} are as in node patterns above; when \emph{spec} is omitted, the abbreviated version can be used. As an example, if we want edges that are either undirected, or directed from right to left, we could write
\pmb{\gqltext{<}}\lstinline{~[e]~}. 

Similarly to node patterns, we can restrict which edges we wish to return by specifying their labels, or using the \gqlkw{WHERE} clause, as was shown earlier with the example that restricted \gqltext{Transfer} edges to those with the amount exceeding 5 million.
Both label expressions and the \gqlkw{WHERE} clause can use the same constructs as in the case of node patterns. Similarly to node patterns, edge patterns need not specify an element variable. This will become useful later on when defining more complex patterns.

\begin{figure}
\begin{tabular}{@{\hspace{0mm}}lcc@{\hspace{0mm}}}
    \toprule
    \textbf{Orientation} & \textbf{Edge pattern} & \textbf{Abbreviation}\\
    \midrule
    Pointing left & \gqlinline!<-[! \emph{spec} \,\gqlinline!]- ! & \gqlinline!<-! \\
    Undirected & \gqlinline! ~[! 
                 \emph{spec} \,\gqlinline!]~ ! & \bf\gqlinline!~! \\
    Pointing right & \gqlinline! -[! \emph{spec} \,\gqlinline!]->! & \gqlinline!->!\\
    Left or undirected & \gqlinline!<~[! \emph{spec} \,\gqlinline!]~ ! &  \gqlinline!<~!\\
    Undirected or right & \gqlinline! ~[! \emph{spec} \,\gqlinline!]~>! &  \gqlinline!~>!\\
    Left or right &  \gqlinline!<-[! \emph{spec} \,\gqlinline!]->! & \gqlinline!<->!\\
    Left, undirected or right & \gqlinline! -[! \emph{spec} \,\gqlinline!]- ! & \bf\gqlinline!-!\\
    \bottomrule
\end{tabular}
\caption{Table of edge patterns.}
\label{fig.edgepatterns}
\end{figure}

\subsection{Building Path Patterns by Concatenation}
Node and edge patterns can be chained together to form a \emph{path pattern}. The most natural way to do this is to ask for edges in the graph, together with their source and target nodes, as in:

\begin{gql}
MATCH (x)-[e]->(y)
\end{gql}

\noindent The bindings generated by this pattern will map 
\gqltext{e} to an edge in the property graph, and \gqltext{x} and \gqltext{y} to its source and target node, respectively. If we do not specify direction and write \gqlinline{(x)-[e]-(y)}, then each edge will be returned twice, once for each direction in which it is traversed.
Edge direction, edge (un)directedness, or any filter on nodes/edges can be used as in the examples above. For instance, to ask for a source account of every transfer that reaches an account with the \gqltext{owner} equal to \gqlinline!'Aretha'!, we write:
\begin{gql}
MATCH (y WHERE y.owner='Aretha')<-[e:Transfer]-(x)
\end{gql}

\noindent Even if the edge pattern is ambiguous about the  orientation of \gqltext{e}, we may wish to refer to this orientation in a postfilter. We can do this using the  predicates \gqltext{e} \gqlkw{IS\:DIRECTED}, 
    \gqltext{s} \gqlkw{IS\:SOURCE\:OF} \gqltext{e}, and \gqltext{d}~\gqlkw{IS\:DESTINATION\:OF} \gqltext{e}. 

One can keep on alternating edge and node patterns to create more complex path patterns.
For instance, the query 

\begin{gql}
MATCH (s)-[e]->(m)-[f]->(t)
\end{gql}

\noindent extracts all directed paths of length two in the graph.
Variable~\gqltext{s} is bound to the id of the source node 
of the path, \gqltext{t} is bound to the id of the target node, and \gqltext{m} to the middle node. Similarly, variables~\gqltext{e} and~\gqltext{f} are bound
to the two edge ids of the path, respectively.
Executed on the graph in Fig.~\ref{fig:propertygraph}, one of the returned bindings is:
\begin{equation*}
          \text{\gqltext{s}}\mapsto \texteltid{a1},
    \quad \text{\gqltext{e}}\mapsto \texteltid{t1},
    \quad \text{\gqltext{m}}\mapsto \texteltid{a3},
    \quad \text{\gqltext{f}}\mapsto \texteltid{t2},
    \quad \text{\gqltext{t}}\mapsto \texteltid{a2}.
\end{equation*}

We can still use the constructs previously described (filters, orientation, labels) 
in each individual edge or node pattern; for example:

\begin{gql}
MATCH (p:Phone WHERE p.isBlocked='yes') 
      ~[e:hasPhone]~(a1:Account)
      -[t:Transfer WHERE t.amount>1M]->(a2)
\end{gql}

\noindent It searches for substantial transfers from accounts into which a login attempt was made from a blocked phone. 
The query still extracts paths of length two, 
but the first edge is undirected and the second is directed and
in forward orientation.
Hence, each returned binding maps \gqltext{p}, \gqltext{a1}, \gqltext{a2} 
to nodes $n_1$, $n_2$, $n_3$,
and \gqltext{e}, \gqltext{t} to edges $e_1$, $e_2$,
such that $e_1$ links $n_2$ and $n_1$, while $e_2$ goes from~$n_2$ to~$n_3$.

Moreover, one may use the same variable multiple times in order to
impose topological constraints on the matched paths via an implicit equi-join on the repeated variable.  
For instance in the query below, the variable \gqltext{s} is used twice; hence this query finds "triangles" of accounts involved in
money transfers; the reuse of variable \gqltext{s} ensures that one starts and ends in the same node:

\begin{gql}
MATCH (s)-[:Transfer]->(s1)-[:Transfer]->(s2)-[:Transfer]->(s)
\end{gql}

Path patterns permit the use of \emph{path variables}: in the returned bindings, such a path variable is
bound to a whole path.
In

\begin{gql}
MATCH p = (s)-[:Transfer]->(s1)-[:Transfer]->(s2)-[:Transfer]->(s)
\end{gql}

\noindent
the variable \gqltext{p} will be bound to paths of length three of \gqltext{Transfer} edges that start and end in the same node.
This variable \gqltext{p} could be returned, or used to compute a more complex value, for instance an aggregate over the bound path. 

As in previous examples, the presence of variables in node or edge patterns is not compulsory.
For instance, the query 

\begin{gql}
MATCH (p:Phone)~[:hasPhone]~(s:Account)-[t:Transfer]->
      (d:Account)~[:hasPhone]~(p)
\end{gql}

\noindent extracts the transfers
between accounts that were accessed from the same phone.
Since the \gqltext{hasPhone} edges have no additional property, there is no need to return them.
It thus returns two bindings:
\begin{gather*}
          \text{\gqltext{p}} \mapsto {\texteltid{p1}},
    \quad \text{\gqltext{s}} \mapsto {\texteltid{a5}}, 
    \quad \text{\gqltext{t}} \mapsto {\texteltid{t8}}, 
    \quad \text{\gqltext{d}} \mapsto {\texteltid{a1}}
    \\
          \text{\gqltext{p}} \mapsto {\texteltid{p2}},
    \quad \text{\gqltext{s}} \mapsto {\texteltid{a3}}, 
    \quad \text{\gqltext{t}} \mapsto {\texteltid{t2}}, 
    \quad \text{\gqltext{d}} \mapsto {\texteltid{a2}}
\end{gather*}

\subsection{Graph Patterns}

Graph patterns combine several path patterns together. As an example, consider 
the fraud detection query looking for substantial transfers from accounts into which a login attempt was made from a blocked phone. It  could be alternatively written as

\begin{gql}
MATCH (p:Phone WHERE p.isBlocked='yes')~[:hasPhone]~(s:Account),
      (s)-[t:Transfer WHERE t.amount>1M]->()
\end{gql}

\noindent
by splitting the path into two edges and reusing the variable \gqltext{s} to indicate that these edges share a node. 
In general, we can put arbitrarily many path patterns together.
For example, we can modify the above query by adding a condition that another login attempt into an account was made that did not use a phone: 

\begin{gql}
MATCH (s:Account)-[:SignInWithIP]-(),
      (s)-[t:Transfer WHERE t.amount>1M]->(),
      (s)~[:hasPhone]~(p:Phone WHERE p.isBlocked='yes')
\end{gql}

\noindent
This pattern looks for three edges coming out of \gqltext{s} which is cumbersome to write as a single path.
In such graph patterns, each path pattern produces a set of mappings, which are then joined using variables that are shared between them. 

\OMIT{
\subsection{Grouping}
Subexpressions within a path pattern may be grouped 
using matching pairs of either parentheses of square 
brackets. In common with many computer languages,
this can be done to override the operator precedence 
shown in Figure \ref{fig.regex}, or simply for clarity.

An important capability of a parenthesized path pattern 
is an optional WHERE clause. We will see examples of 
this when we consider quantification, next.
}

\subsection{Quantifiers and Group Variables}
GPML includes quantifiers similar 
to those in Perl and other common ``regex'' tools. 
Figure \ref{fig.quantifiers} lists the quantifiers of GPML.
Quantifiers are written as postfix operators on either 
a single edge pattern or a parenthesized path pattern 
(a path pattern enclosed in parentheses or square brackets with an optional \gqlkw{WHERE} clause). 
For example, a path of length 2 to 5 of \gqltext{Transfer} edges can be sought as follows:

\begin{gql}
MATCH (a:Account)-[:Transfer]->{2,5}(b:Account)
\end{gql}

As an example using a parenthesized path pattern, consider the problem of finding paths of~2 to~5 \gqltext{Transfer} edges between accounts with the same owner:

\begin{gql}
MATCH [(a:Account)-[:Transfer]->(b:Account) WHERE a.owner=b.owner]{2,5}
\end{gql}

The pattern does not provide
a strict alternation of node and edge patterns; it will be unrolled two to five times, to obtain a sequence of bindings of variables \gqltext{a b a b}\dots The \gqlkw{WHERE} clause here applies to each such pair of \gqltext{a} and \gqltext{b} bindings separately. 
As variables \gqltext{a} and \gqltext{b} will occur multiple times in such a sequence, 
the notion of binding an element variable to a unique
node or vertex is insufficient in this case, and 
must be expanded to embrace \emph{path bindings}
in which each variable in that sequence is mapped to a 
graph element. 
In addition, at the transition between groups, the previous binding of \gqltext{b} must be the same as the next binding of \gqltext{a}. More about the exact mechanism of such bindings will follow from the detailed example in Section \ref{sec:detailedexample}.

As for a quantifier on a bare edge pattern, this is 
understood by supplying anonymous node patterns to
its left and right. For example, if we are interested in pairs of accounts involved in a chain of large transfers of length between 2 and 5 we could write

\begin{gql}
MATCH (a:Account) 
      [()-[t:Transfer]->() WHERE t.amount>1M]{2,5} 
      (b:Account)
\end{gql}

\begin{figure}
\begin{tabular}{cl}
\toprule
    \textbf{Quantifier} & \textbf{Description} \\
\midrule
    \gqlinline!{m,n}! & between \gqlinline!m! and \gqlinline!n! repetitions
    \\
    \gqlinline!{m,}! & \gqlinline!m! or more repetitions 
    \\
    \gqlinline!*! & equivalent to \gqlinline!{0,}!
    \\
    \gqlinline!+! & equivalent to \gqlinline!{1,}!
    \\
\bottomrule
\end{tabular}
\vspace{-3mm}
\caption{Table of quantifiers}
\label{fig.quantifiers}
\end{figure}

Variable references are categorized as either \emph{singleton} or \emph{group}. 
For example, in predicates such as \gqlinline!x.isBlocked='yes'!, the variable \gqltext{x} is referenced as a singleton.
Intuitively, a reference is group if you have to cross a quantifier to get from the reference to the declaration of the variable. 
To explain this, consider a modification of the above example where we are only interested in chains of transfers with the total value over 10 million: 

\begin{gql}
MATCH (a:Account) 
      [()-[t:Transfer]->() WHERE t.amount>1M]{2,5} 
      (b:Account)
WHERE SUM(t.amount)>10M
\end{gql}

\noindent In this example, the predicate in the edge pattern
references \gqltext{t} as a singleton, since one does not have to
cross the quantifier \gqlinline!{2,5}! to reach its declaration
which is in the same edge pattern.
The predicate in the final \gqlkw{WHERE} clause, used here as a postfilter, references \gqltext{t} in the aggregate \gqlkw{SUM}. 
In this location one has to cross the quantifier to reach the declaration of \gqltext{t}, making it a group reference.

\subsection{Path Pattern Union \& Multiset Alternation}

In GPML there are two forms of union, called path pattern union (with set semantics) and multiset alternation (with multiset semantics). Path pattern union is indicated by an infix vertical bar, whereas multiset alternation uses \gqltext{|+|} as its infix operator.
An example of path pattern union is

\begin{gql}
MATCH (c:City) | (c:Country)
\end{gql}

\noindent In the sample graph, there are two \gqltext{Country} nodes
(\texteltid{c1} and \texteltid{c2}) and one \gqltext{City} node (\texteltid{c2}). Thus the first
operand produces two results  $\text{\gqltext{c}}\mapsto {\texteltid{c1}}$ and $\text{\gqltext{c}}\mapsto {\texteltid{c2}}$ 
and the second operand produces the single result $\text{\gqltext{c}}\mapsto {\texteltid{c2}}$.
%
Thus the operands produce a duplicate binding
to \texteltid{c2}, which will be reduced to a single solution in 
the final result, which has one binding to \texteltid{c1} and one
binding to \texteltid{c2}.

Rewriting this example with multiset alternation, we have

\begin{gql}
MATCH (c:City) |+| (c:Country)
\end{gql}

\noindent This pattern returns \emph{three} results, 
one result binding \texteltid{c1} and two results binding \texteltid{c2}.

Another example of deduplication using path pattern union
is overlapping quantifiers, as in this example:

\begin{gql}
MATCH ->{1,5} | ->{3,7}
\end{gql}

\noindent The two quantifiers overlap between 3 and 5; 
consequently when the results are deduplicated the query
is equivalent to:

\begin{gql}
MATCH ->{1,7}
\end{gql}

\noindent
Using multiset alternation would not deduplicate the overlap in the quantifiers. 

\OMIT{ 
It is comparatively easy to see how to 
deduplicate these examples at compile time. However, a 
general algorithm to simplify arbitrary path pattern
unions in the complete language, including \gqlinline!WHERE! clauses,
restrictors, nested quantifiers, etc., is not so clear.
In addition, even if a transform is known, 
an implementer may judge that making such
transforms for patterns that their users do not write
frequently is not worth the implementation cost. Performing
deduplication at run-time is also expensive, and from 
a user perspective, it may be acceptable to get a result
that is a multiset, rather than waiting a long time for a set result. (SQL users are known to accept \gqlkw{UNION}~\gqlkw{ALL} for
its speed over \gqlkw{UNION}~\gqlkw{DISTINCT} for its set semantics.)
For these reasons GPML provides both path pattern union
and multiset alternation. 
}

While this particular example is easy to 
deduplicate at compile time, a general  
algorithm to simplify arbitrary path pattern
unions in the language that includes \gqlinline!WHERE! clauses,
restrictors, nested quantifiers, etc., is not so clear and is likely to have a very high complexity, making it an unattractive option for the 
implementer. 
Run-time deduplication is also expensive, and from 
a user perspective, it may be acceptable to get a result
that is a multiset, rather than waiting a long time for a set result. This is akin to the SQL 
situation with \gqlkw{UNION}~\gqlkw{ALL} chosen for
its speed over \gqlkw{UNION}~\gqlkw{DISTINCT} guaranteeing set semantics.
For these reasons GPML provides both path pattern union
and multiset alternation. 

\subsection{Conditional Variables}
Consider the following query.

\begin{gql}
MATCH [(x)->(y)] | [(x)->(z)]
\end{gql}

\noindent Node variable \gqltext{x} will be bound if either operand
of the path pattern union binds. Variables \gqltext{y} and \gqltext{z}, on 
the other hand, are only bound by one operand but not 
by the other. We say that \gqltext{x}  is an 
\emph{unconditional singleton} whereas \gqltext{y}  and \gqltext{z} 
are \emph{conditional singletons}.

Implicit equi-joins on conditional 
singletons are disallowed, because they lack intuitive semantics.  For instance, in the illegal query

\begin{gql}
MATCH [(x)->(y)] | [(x)->(z)], (y)->(w)
\end{gql}

\noindent 
does the fact that \gqltext{y} must bind in  the second path pattern imply that \gqltext{y} must bind in the path pattern union, thereby effectively excluding the results of the second operand?
Implicit equi-joins on conditional singletons were forbidden in order to eliminate such doubts.

Conditional singletons are also introduced by 
the question mark operator, a postfix operator similar
to the quantifiers; for example

\begin{gql}
MATCH (x) [->(y)]?
\end{gql}

\noindent In most ``regex'' tools, 
the postfix question mark operator
is equivalent to the quantifier \gqlinline!{0,1}!. 
In GPML the two operators have almost the same 
semantics. The difference is that the quantifier \gqlinline!{0,1}!
exposes all variables as group, whereas the question
mark operator exposes singletons as conditional singletons.
This distinction was made so that query-generating 
tools can solicit upper and lower bounds
for a quantifier from the user without having to watch
for \gqlinline!{0,1}! as a special case. 



To provide an example of the use of conditional variables in our running example, consider a query asking for 
accounts from which mobney was transferred to either a blocked account
or an account with a login from a blocked phone. It could be written as path patetrn union:

\begin{gql}
MATCH [(x:Account)-[:Transfer]->(y:Account WHERE y.isBlocked='yes')] |
      [(x:Account)-[:Transfer]->()-[:hasPhone]-(p WHERE p.isBlocked='yes')]
\end{gql}

Alternatively, one can use the question mark operator:

\begin{gql}
MATCH (x:Account)-[:Transfer]->(y:Account) [-(:hasPhone)-(p)]?
WHERE y.isBlocked='yes' OR p.isBlocked='yes'
\end{gql}

If the optional part of the pattern is not matched, then the condition \gqltext{p.isBlocked='yes'} is not true, and thus the account must be blocked to ensure that the condition in \gqlkw{WHERE} is true. 



\subsection{Graphical Predicates}
In addition to familiar predicates such as comparison
(e.g., equal to, greater than, etc.) and \gqlkw{IS}~\gqlkw{NULL}, GPML
specifies some predicates relevant to graphs (applicable to singleton variable references).


Many of the edge patterns are ambiguous about the 
orientation of an edge. For example, \gqltext{-[e]-}
matches a directed edge pointing either left or right,
or an undirected edge. The user may wish to interrogate
the orientation of the edge bound to \gqltext{e}. For this, GPML
provides the following predicates:
\begin{description}
    \item[\gqltext{e\:IS\:DIRECTED}] is true if \gqltext{e} is bound to 
    a directed edge;
    \item[\gqltext{s\:IS\:SOURCE\:OF\:e}] is true if \gqltext{s}  is bound to
    the source of \gqltext{e};
    \item[\gqltext{d\:IS\:DESTINATION\:OF\:e}] is true if \gqltext{d}  is 
    bound to the target of \gqltext{e}. 
\end{description}

SQL/PGQ is not able to use the equals sign to test for
equality of element references, because \gqltext{p=q}
is interpreted as an equality test on two column values,
therefore \gqltext{p}  and \gqltext{q}  cannot be interpreted as element
references. GQL, on the other hand, will permit such equality tests of
element references. To provide a portable way to 
test equality of element references, GPML provides the 
following predicates:

\begin{itemize}
    \item \gqltext{SAME(p, q, ...)}: test whether all 
    element references in the list are bound to the 
    same graph element. 
    \item \gqltext{ALL\_DIFFERENT(p, q, ...)}: test 
    whether the element references in the list are 
    pairwise distinct.
\end{itemize}
\noindent The element references listed in \gqltext{SAME} or
\gqltext{ALL\_DIFFERENT} must be unconditional singletons
because the semantics of such expressions for 
conditional singletons would be ambiguous. 

\section{Assuring termination}\label{sec:sr}
Written without any restrictions, GPML queries may not terminate as they will return infinitely many matches. Consider for example, 

\begin{gql}
MATCH p = (a WHERE a.owner='Dave')-[t:Transfer]->*
          (b WHERE b.owner='Aretha')
\end{gql}

\noindent It asks for paths with any number of transfers between the account owned by \gqltext{Dave} and the account owned by \gqltext{Aretha}. If evaluated over the graph of Figure \ref{fig:propertygraph}, this query  would have an infinite number of matches, since the path matched by \gqltext{p} could include a \gqltext{Transfer} loop
(e.g., $\pathvalue{a3,t7,a5,t8,a1,t1,a3}$) any number of times. 

To prevent this behaviour, GPML queries must {\em demonstrably terminate}; in particular,
the number of matches must be finite. To achieve this, GPML uses \emph{restrictors} and \emph{selectors}. Every unbounded quantifier (such as \gqlinline!*! above) must be contained in the scope of either a restrictor or a selector or both. 

\subsection{Restrictors and Selectors}

\paragraph*{Restrictors.} 
A restrictor is a path predicate (that is, it imposes a condition stating which paths are acceptable) such that the number of matches cannot be infinite.
For instance, \gqlkw{TRAIL} forbids the matched path to repeat edges; there are only finitely many edges, therefore there are only finitely many paths with no repeated edges in a graph. Restrictors are listed in Figure \ref{fig.restrictors}.
Restrictors may be placed either at the head of a path
pattern, or at the head of a parenthesized path pattern. 
If the above example is written as 

\begin{gql}
MATCH TRAIL p = (a WHERE a.owner='Dave')-[t:Transfer]->*
                (b WHERE b.owner='Aretha')
\end{gql}

\noindent then, executed on the graph of Fig.~\ref{fig:propertygraph}, returns 
three bindings for \gqltext{p}: 
\begin{gather*}
    \pathvalue{a6,t5,a3,t2,a2}\\
    \pathvalue{a6,t6,a5,t8,a1,t1,a3,t2,a2}\\
    \pathvalue{a6,t5,a3,t7,a5,t8,a1,t1,a3,t2,a2}
\end{gather*}
Note that the last path repeats the node \texteltid{a3}; it is allowed by \gqlkw{TRAIL}
but would be forbidden by the restrictor \gqlkw{ACYCLIC}.
Note also that 
\begin{equation*}
\pathvalue{a6,t5,a3,t2,a2,t3,a4,t4,a6,t5,a3,t2,a2}
\end{equation*}
\noindent which traverses the \gqltext{Transfer} cycle starting and ending in \texteltid{a6}, is not a trail, and is thus not returned.

\paragraph*{Selectors.} 
A selector is an algorithm that 
conceptually partitions the solution space on the endpoints 
and selects a finite set of matches from each partition. 
For instance, \gqlkw{ALL}~\gqlkw{SHORTEST} keeps all paths 
having the same shortest length within each partition 
defined by a pair of endpoints for which any solution 
exists. Note that the shortest length can differ from 
partition to partition. Selectors are listed in 
Figure \ref{fig.selectors}. Selectors may only be placed
at the head of a path pattern.
As an example, 
applying a selector to the example from the beginning of Section \ref{sec:sr} can be done as follows:

\begin{gql}
MATCH ANY SHORTEST 
  p = (a WHERE a.owner='Dave')-[t:Transfer]->*(b WHERE b.owner='Aretha')
\end{gql}

\noindent Here we wish only one of the shortest paths between the nodes \texteltid{a6} and \texteltid{a2}. 
In this case, there is  only one shortest path between these nodes and thus \gqltext{p} is bound to $\pathvalue{a6,t5,a3,t2,a2}$.

\paragraph*{Combining restrictors and selectors.}
At the conceptual level, restrictors can be seen as operating during pattern matching 
while selectors operate afterwards. 
\footnote{This is of course 
only an intuitive description: in the case of selectors, 
the engine will not compute an infinite number of matches 
and then only keep some. However, to make the specification 
declarative ("tell me what you want, not how to get it")
the specification indeed describes how the 
infinite set is defined and then which finite selection from it should be returned.}
That is, if combined, selectors are always applied \emph{after} restrictors. For instance, consider the query:

\begin{gql}
MATCH ALL SHORTEST TRAIL
  p = (a WHERE a.owner='Dave')-[t:Transfer]->*
      (b WHERE b.owner='Aretha')-[r:Transfer]->*(c WHERE c.owner='Mike')
\end{gql}

\noindent It selects the shortest paths \emph{among} the trails going from node \texteltid{a6} to node~\texteltid{a3} and passing through~\texteltid{a2}.
It returns two bindings for \gqltext{p}:
\begin{gather*}
    \pathvalue{a6,t5,a3,t2,a2,t3,a4,t4,a6,t6,a5,t8,a1,t1,a3} \\
    \pathvalue{a6,t6,a5,t8,a1,t1,a3,t2,a2,t3,a4,t4,a6,t5,a3}
\end{gather*}
The path $\pathvalue{a6,t5,a3,t2,a2,t3,a4,t4,a6,t5,a3}$
is not considered: it is shorter but it is not a trail.


This difference has one notable consequence. 
Consider a query $Q$ with no selector or restrictor, and assume that~$Q$ has matches.
Then, adding a selector to~$Q$ might reduce the number of matches, but the resulting query will always have at least one match.
On the other hand, adding a restrictor to~$Q$ might yield a query with no matches at all. 

For example, consider the query

\begin{gql}
MATCH (p:Account WHERE p.owner='Natalia')->{1,10}
      (q:Account WHERE q.owner='Mike')->{1,10}
      (r:Account WHERE r.owner='Scott')
\end{gql}

\noindent In the example graph, 
$\pathvalue{a5,t8,a1,t1,a3,t7,a5,t8,a1}$
is a solution to this query, and in fact it is a shortest
solution in the partition defined by the endpoints 
(a5,a1). However, it repeats the edge t8, so it fails
the restrictor \gqlkw{TRAIL} (as well as 
\gqlkw{SIMPLE} and \gqlkw{ACYCLIC}. Thus adding a selector
such as \gqlkw{ALL SHORTEST} will still have a result, whereas
adding a restrictor such as \gqlkw{TRAIL} will have no result. 


\begin{figure}
\renewcommand{\arraystretch}{1}
\begin{tabularx}{\linewidth}{p{.22\linewidth}X}
\toprule
    \textbf{Keyword} & \textbf{Description} \\
\midrule
    \gqlkw{TRAIL} & No repeated edges.
    \\
    \gqlkw{ACYCLIC} & No repeated nodes.
    \\
    \gqlkw{SIMPLE} & No repeated nodes, except that the first and last nodes may be the same.
    \\
\bottomrule
\end{tabularx}
\caption{Table of restrictors}
\label{fig.restrictors}
\end{figure}

\begin{figure}
\renewcommand{\arraystretch}{1.3}
\begin{tabularx}{\linewidth}{>{\raggedright}p{.228\linewidth}X}
\toprule
    \textbf{Keyword} & \textbf{Description} \\
\midrule
    \gqlkw{ANY} \gqlkw{SHORTEST} & Selects one path with shortest length from each partition.
    Non-deterministic. 
    \\
    \gqlkw{ALL} \gqlkw{SHORTEST} & Selects all paths in each partition that have the minimal length in the partition. Deterministic. 
    \\
    \gqlkw{ANY} & Selects one path in each partition arbitrarily. 
    Non-deterministic.
    \\
    \gqlkw{ANY} $k$ & Selects arbitrary $k$ paths in each partition (if fewer than~$k$, then all are retained). 
    Non-deterministic.
    \\
   \gqlkw{SHORTEST} $k$  & Selects the shortest $k$ paths (if fewer than~$k$, then all are retained).  
    Non-deterministic.
    \\
    \gqlkw{SHORTEST} $k$ \gqlkw{GROUP} &  
    Partitions by endpoints, sorts each partition by path length, groups paths with the same length, then selects all paths in the first $k$ groups from each partition
    (if fewer than~$k$, then all are retained). Deterministic.
    \\
\bottomrule
\end{tabularx}
\caption{Table of selectors}
\label{fig.selectors}
\end{figure}

\subsection{Prefilters and Postfilters of Selectors}
When working with selectors, it is important to 
differentiate \emph{prefilters} from \emph{postfilters}. 
A prefilter is a predicate applied before selection;
a postfilter is a predicate applied after selection.
In GPML, postfilters are expressed in the final \gqlkw{WHERE}
clause, whereas prefilters are expressed in element
patterns or parenthesized path patterns within the 
selector's path pattern. 

For example, suppose we want to find the shortest path
from Scott to Charles passing through an account that is 
blocked:

\begin{gql}
MATCH ALL SHORTEST (p:Account WHERE p.owner='Scott')->+ 
                   (q:Account WHERE q.isBlocked='yes')->+
                   (r:Account WHERE r.owner='Charles')
\end{gql}

\noindent Note that all predicates in this example 
are expressed
in element patterns, therefore they are all prefilters.
The only solution is the $\pathvalue{a1,t1,a3,t2,a2,t3,a4,
t4,a6,t5,a3,t7,a5}$, in which \gqltext{q} is bound to \texteltid{a4} (\gqltext{Jay},
the only blocked account).

It would be a mistake to place the predicate on \gqltext{q} in the final \gqlkw{WHERE} clause, like this:

\begin{gql}
MATCH ALL SHORTEST (p:Account WHERE p.owner='Scott')->+
                   (q:Account)->+
                   (r:Account WHERE r.owner='Charles')
WHERE q.isblocked='yes'
\end{gql}

\noindent The shortest 
path from Scott to Charles is $\pathvalue{a1,t1,a3,t7,a5}$
with \gqltext{q} bound to \texteltid{a3},
which will be the result of the selector, but this
result is then filtered out by the final \gqlkw{WHERE} clause
because \texteltid{a3} is not blocked. Consequently this
query finds no result. The original problem statement
had the predicate 
\emph{``passing through an \gqltext{Account} that is blocked''}
as a prefilter, therefore the query placing it 
as a postfilter is incorrect.

\OMIT{
\subsection{Independence of Selectors}
As mentioned, a selector can only be expressed at the 
head of a path pattern. If a graph pattern contains
two or more selectors, they must be in separate 
path patterns. GPML requires that multiple selectors
in the same graph pattern must be independent. 

Independence of selectors is defined as follows. An 
unconditional singleton node variable that is the first or 
the last element pattern in a selector is called a 
\emph{boundary variable}. All other element variables
declared within a selector are called 
\emph{strict interior variables}. 

Note that the selectors are defined by partitioning 
the solution space on the bindings to the boundary
variables. This means that the bindings to the strict 
interior variables depend on the bindings to the 
boundary variables. Independence of selectors means that
one selector may not reference or form an implicit
equijoin to a strict interior variable of another
selector. This consideration gives the following rules for 
independence of selectors:
\begin{itemize}
    \item A strict interior variable of one selector may not be equivalent to either a boundary variable or strict
    interior variable of another selector.
    \item A predicate in one selector may not reference
    a strict interior variable of another selector.
\end{itemize}
}

\subsection{Aggregates of Unbounded Variables}
There is another subtle way of having non-terminating queries that must be ruled out by GPML. Consider this query:

\begin{gql}
MATCH ALL SHORTEST [ (x)-[e]->*(y) WHERE COUNT(e.*)/(COUNT(e.*)+1)>1 ]
\end{gql}

Note carefully that the predicate is within a parenthesized
path pattern, therefore it is a prefilter. As a prefilter,
the group variable \gqltext{e} has not yet passed through the 
selector \gqlkw{ALL SHORTEST}; consequently the 
predicate sees \gqltext{e} as effectively unbounded.

If a match has 0 edges, the quotient in the \gqlkw{WHERE} clause
is 0; if it has 1 edge, the quotient is 1/2; if it has 
2 edges, the quotient is 2/3; etc. The quotient can 
never exceed 1. This is easily seen by human reason,
but generalizing this observation to any aggregate on
an unbounded element reference is not obvious.
For example, suppose the aggregate were \gqlkw{AVG}\gqltext{(e.x)};
the behavior of \gqlkw{AVG} on an unbounded collection of
property references is not easy to anticipate.
A few aggregates (\gqlkw{MAX}, \gqlkw{MIN}, \gqlkw{COUNT}) are monotonic,
which might permit reasoning on simple expressions 
such as linear combinations, but it is currently
expected that GPML will simply prohibit all predicates
on unbounded groups to ensure termination.

The query can be made acceptable -- i.e., turned into a terminating one -- in several ways.
First, the predicate might be converted to a postfilter,
like this:

\begin{gql}
MATCH ALL SHORTEST (x)-[e]->*(y)
WHERE COUNT(e.*)/(COUNT(e.*)+1) > 1 
\end{gql}

Note that all we did was remove the square brackets,
which moved the predicate out of the path pattern and
into the final \gqlkw{WHERE} clause. This made \gqltext{e}
effectively bounded. Of course any results produced by the selector will
be filtered out by the postfilter; therefore the result
of this query is empty. However, turining the predicate into a postfilter and using a selector first ensures the finiteness of the result, and thus termination. 

A different approach is to retain the prefilter 
but insure that \gqltext{e}is effectively bounded 
within the selector. Certainly changing the quantifier
to have a static upper bound such as \{0,10\} would do. 
Alternatively a restrictor could be used, like this:

\begin{gql}
MATCH ALL SHORTEST [ TRAIL (x)-[e]->*(y) WHERE COUNT(e.*)/(COUNT(e.*)+1) > 1 ]
\end{gql}

Reasoning about this pattern we can see that the result will be
empty. An implementation is of course free to 
implement algorithms that detect some queries that are 
unsatisfiable and thereby terminate quickly. 
Alternatively, the implementation could actually 
tackle the query as written and eventually arrive
at the conclusion that the result set is empty.


\section{Execution model by example}\label{sec:detailedexample}
\newcounter{anoncounter}
\newcommand{\rvarnode}[2][]{{\square}_{#2}^{#1}}
\newcommand{\rvaredge}[2][]{{-}_{#2}^{#1}}
\newcommand{\varnode}[2][]{\setcounter{anoncounter}{#2}\rvarnode[#1]{\roman{anoncounter}}}
\newcommand{\varedge}[2][]{\setcounter{anoncounter}{#2}\rvaredge[#1]{\roman{anoncounter}}}
\newcommand{\varb}[1]{{\text{\gqltext{b}}^{#1}}}

We illustrate GPML evaluation using a step-by-step example. The output of the example query is a set, or a multiset, of {\em reduced path bindings}. A {\em path binding} is a sequence of \emph{elementary bindings}, which in turn are pairs of a variable and a graph element. 
A \emph{variable} is either a variable that occurs in the pattern, or a new variable corresponding to an anonymous node or edge. 
Each variable, anonymous or not, can be further indexed if it appears under the scope of a quantifier: the index indicates the iteration in which it appears. For the same variable \gqltext{a} under the scope of a quantifier \verb+{1,}+ we can have variables $\gqltext{a}^1,\gqltext{a}^2,\gqltext{a}^3,\ldots$ that correspond to different iterations of the quantifier. 
Obtained path bindings are then {\em reduced}, stripping all annotations, and {\em deduplicated}, to remove duplicates. If multiple path patterns appear after \gqlkw{MATCH}, the process is repeated for all of them, and the result is joined on identical singleton variables, and possibly further filtered if there is a \gqlkw{WHERE} clause. Selectors and restrictors are used to ensure that the output is finite in the presence of quantifiers.

The key steps in the pattern matching execution model, as reflected in the forthcoming standard, are as follows.

\begin{description}
\item[Normalization]\
GPML provides syntactic sugar to help write patterns; this step puts 
patterns in a canonical form.
    
\item[Expansion] The pattern is expanded into a 
set of \emph{rigid patterns} without any kind of disjunction.
Intuitively, a rigid pattern is one that could be expressed by an SQL equi-join query.
Formally, it is a pattern without quantifiers, union or multiset alternation. 
The expansion also annotates each rigid pattern to enable tracking the provenance of the syntax constructs.

\item[Rigid-pattern matching]\quad 
For each rigid pattern, one computes a set of \emph{path bindings}.
Each elementary construct of the rigid pattern is computed independently and then the results are joined together on variables with the same name.

\item[Reduction and deduplication]\quad
The path bindings matched by the rigid patterns are \emph{reduced} by removing annotations. The reduced path bindings are then collected into a set.
This implies a \emph{deduplication} step since different path bindings might become equal after reduction, and only one copy is kept.
\end{description}

\subsection{Running Example} 

The remainder of this section gives a detailed account of how pattern matching is computed for the following query.

\begin{gql}
MATCH TRAIL (a WHERE a.owner='Jay')
            [-[b:Transfer WHERE b.amount>5M]->]+ 
            (a) [-[:isLocatedIn]->(c:City) |
                 -[:isLocatedIn]->(c:Country)]
\end{gql}

This query finds sequences of transfers of arbitrary length that start and end with account owner \gqltext{Jay}, as well as the location (city or country) of \gqltext{Jay}. To ensure termination, the \gqlkw{TRAIL} mode is used.  

\subsection{Normalisation}

The first step of normalisation makes each sequence of
node and edge patterns consistent.
More precisely, each such sequence must start and end with a node pattern; and alternate between node and edge patterns.
In addition, syntactic sugar is expanded, e.g., quantifier \texttt{+} is replaced by \texttt{\{1,\}}.
Hence, the pattern becomes:

\begin{gql}
(a WHERE a.owner='Jay')  
[()-[b:Transfer WHERE b.amount>5M ]->()]{1,} 
(a) [()-[:isLocatedIn]->(c:City) | 
     ()-[:isLocatedIn]->(c:Country)]
\end{gql}

\noindent We then introduce a fresh variable into each anonymous node and edge pattern (that is, a pattern that is not assigned to a variable). 
The fresh node and edge variables are denoted by~$\rvarnode{x}$ 
and $\rvaredge{x}$, respectively, for some index~$x$. The pattern then becomes

\noindent\hspace*{-1mm}\begin{gql}
(a WHERE a.owner='Jay')  
[($\varnode{1}$)-[b:Transfer WHERE b.amount>5M]->($\varnode{2}$)]{1,}
(a)[[($\varnode{3}$)-[$\varedge{1}$:isLocatedIn]->(c:City)] |
    [($\varnode{4}$)-[$\varedge{2}$:isLocatedIn]->(c:Country)]]
\end{gql}

\subsection{Expansion}

Expansion turns the pattern into a set of rigid patterns, which fix the
number of iterations for each quantifier and a disjunct for each union or alternation.
In our query, this amounts to choosing a disjunct in the union \gqlinline{|}, and expanding the  quantifier \gqlinline!{1,}!.
While unbounded quantifiers such as \gqlinline!{1,}! make the set infinite,  techniques of Section~\ref{sec:sr} will make the evaluation feasible. 
The following pattern is one possible expansion, where we  expanded the quantifier once and chose the left side of the path pattern union: 

\begin{gql}
(a WHERE a.owner='Jay')   
($\varnode[1]{1}$)-[$\varb{1}$:Transfer WHERE $\varb{1}$.amount>5M]->($\varnode[1]{2}$) 
(a) ($\varnode{3}$)-[$\varedge{1}$:isLocatedIn]->(c:City)
\end{gql}

In general, one such pattern is obtained for
each~$n\in\mathbb{N}\setminus\{0\}$ 
and each~$\ell\in\{\gqltext{City},\gqltext{Country}\}$: we expand the quantifier $n$ times, and choose either the \gqltext{City} or the \gqltext{Country} option of the path pattern union. Such a 
pattern is denoted by $\theta_{n,\ell}$. As an example, the pattern below is 
$\theta_{n,\gqltext{City}}$. 

\begin{gql}
(a WHERE a.owner='Jay')  
($\varnode[1]{1}$)-[$\varb{1}$:Transfer WHERE $\varb{1}$.amount>5M]->($\varnode[1]{2}$) 
($\varnode[2]{1}$)-[$\varb{2}$:Transfer WHERE $\varb{2}$.amount>5M]->($\varnode[2]{2}$) 
     $\raisebox{-0.5mm}{\smash{\vdots}}$              $\raisebox{-0.5mm}{\smash{\vdots}}$              $\raisebox{-0.5mm}{\smash{\vdots}}$
($\varnode[n]{1}$)-[$\varb{n}$:Transfer WHERE $\varb{n}$.amount>5M]->($\varnode[n]{2}$) 
(a) ($\varnode{3}$)-[$\varedge{1}$:isLocatedIn]->(c:City) 
\end{gql}

\noindent
The prototype $\theta_{n,\mathtt{Country}}$ is similar, only the last line changes.

Each group variable is marked by a superscript, corresponding to the iteration of the quantifier the variable is in.

The next step is a clean-up: every node pattern with an anonymous variable is deleted 
if it is adjacent to another node pattern.
The resulting pattern is denoted by $\pi_{n,\ell}$. 
For instance, $\pi_{n,\gqltext{City}}$ is:

\begin{gql}
(a WHERE a.owner='Jay') 
-[$\varb{1}$:Transfer WHERE $\varb{1}$.amount>5M]->($\varnode[1]{2}$) 
-[$\varb{2}$:Transfer WHERE $\varb{2}$.amount>5M]->($\varnode[2]{2}$) 
     $\raisebox{-0.5mm}{\smash{\vdots}}$              $\raisebox{-0.5mm}{\smash{\vdots}}$              $\raisebox{-0.5mm}{\smash{\vdots}}$
-[$\varb{n-1}$:Transfer WHERE $\varb{n-1}$.amount>5M]->($\varnode[n-1]{2}$) 
-[$\varb{n}$:Transfer WHERE $\varb{n}$.amount>5M]->(a)
-[$\varedge{1}$:isLocatedIn]->(c:City) 
\end{gql}


\subsection{Computation of Path Binding}
\newenvironment{pathbinding}[1][cccccccccccccccccccccccccccccccc]{\begin{center} \renewcommand{\tabcolsep}{2pt}\begin{tabular}{#1}}{\end{tabular}\end{center}}

The matches of rigid patterns are computed independently.
The result of the computation is called a \emph{path binding}, 
which in turn is a sequence of \emph{elementary bindings}. An
\emph{elementary binding} is a pair of a variable and a graph element. It is convenient to portray these pairs as tables with two rows: the first row contains the variables, the second row contains the graph elements, and each column is an elementary binding. 
An example path binding is 
\begin{pathbinding}
\gqltext{a} & $\varb{1}$ & $\varnode[1]{2}$ \\
\texteltid{a4} & \texteltid{t4} & \texteltid{a6} \\
\end{pathbinding}

Consider the rigid path pattern~$\pi_{4,\gqltext{City}}$. 
Each node-edge-node pattern is computed independently using the input graph,
and then equi-joined on variables with the same name.
The first node-edge-node part of~$\pi_{4,\gqltext{City}}$ is 

\begin{gql}
(a WHERE a.owner='Jay')-[$\varb{1}$:Transfer WHERE $\varb{1}$.amount>5M]->($\varnode[1]{2}$)
\end{gql}

\noindent In the graph it matches only one path binding, the one shown above.  The second node-edge-node part is 

\begin{gql}
($\varnode[1]{2}$)-[$\varb{2}$:Transfer WHERE $\varb{2}$.amount>5M]->($\varnode[2]{2}$)
\end{gql}

\noindent and the path bindings it matches are given in the left column below. The path bindings matched by the third, fourth and fifth parts of~$\pi_{4,\gqltext{City}}$ are given in the three other columns.
\begin{pathbinding}
$\varnode[1]{2}$ & $\varb{2}$ & $\varnode[2]{2}$
&\hspace*{15mm}&
$\varnode[2]{2}$ & $\varb{2}$ & $\varnode[3]{2}$
&\hspace*{15mm}&
$\varnode[3]{2}$ & $\varb{4}$ & \gqltext{a} 
&\hspace*{15mm}&
\gqltext{a} & $\varedge{1}$ & \gqltext{c}
\\
\texteltid{a6} & \texteltid{t5} & \texteltid{a3} 
&&
\texteltid{a6} & \texteltid{t5} & \texteltid{a3}
&&
\texteltid{a6} & \texteltid{t5} & \texteltid{a3}
&&
\texteltid{a4} & \texteltid{li4} & \texteltid{c2}
\\
\texteltid{a3} & \texteltid{t2} & \texteltid{a2}
&&
\texteltid{a3} & \texteltid{t2} & \texteltid{a2}
&&
\texteltid{a3} & \texteltid{t2} & \texteltid{a2}
&&
\texteltid{a6} & \texteltid{li6} & \texteltid{c2}
\\
\texteltid{a2} & \texteltid{t3} & \texteltid{a4}
&&
\texteltid{a2} & \texteltid{t3} & \texteltid{a4}
&&
\texteltid{a2} & \texteltid{t3} & \texteltid{a4}
&&
\texteltid{a3} & \texteltid{li3} & \texteltid{c1}
\\
\texteltid{a2} & \texteltid{t3} & \texteltid{a4} 
&&
\texteltid{a2} & \texteltid{t3} & \texteltid{a4}
&&
\texteltid{a2} & \texteltid{t3} & \texteltid{a4} 
&&
\texteltid{a2} & \texteltid{li2} & \texteltid{c2}
\\
\texteltid{a3} & \texteltid{t7} & \texteltid{a5}
&&
\texteltid{a3} & \texteltid{t7} & \texteltid{a5}
&&
\texteltid{a3} & \texteltid{t7} & \texteltid{a5}
&&
\texteltid{a5} & \texteltid{li5} & \texteltid{c1}
\\
\texteltid{a5} & \texteltid{t8} & \texteltid{a1} 
&&
\texteltid{a5} & \texteltid{t8} & \texteltid{a1} 
&&
\texteltid{a5} & \texteltid{t8} & \texteltid{a1} 
&&
\texteltid{a1} & \texteltid{li1} & \texteltid{c1}
\\
\texteltid{a1} & \texteltid{t1} & \texteltid{a3} 
&&
\texteltid{a1} & \texteltid{t1} & \texteltid{a3} 
&&
\texteltid{a1} & \texteltid{t1} & \texteltid{a3} 
&&

\end{pathbinding}
\noindent Note that the matches for each part (i.e., each column above) is computed independently.

The labels are matched and the \gqlkw{WHERE} clauses are checked at this stage. Thus, the edge (\texteltid{a6},\texteltid{t6},\texteltid{a5}) does not appear anywhere above as it fails the \gqlkw{WHERE} condition,  nor does the edge (\texteltid{ip1},\texteltid{sip1},\texteltid{a1}) since it has neither the \gqltext{Transfer} nor the \gqltext{isLocatedIn} label.

Then the path bindings are concatenated by an implicit equi-join on variables with the same name. In the end there is only one path binding for~$\pi_{4,\gqltext{City}}$ given below.
\begin{pathbinding}
  \gqltext{a} 
  & $\varb{1}$ & $\varnode[1]{2}$ 
  & $\varb{2}$ & $\varnode[2]{2}$
  & $\varb{3}$ & $\varnode[3]{2}$
  & $\varb{4}$ & \gqltext{a} & $\varedge{1}$ & \gqltext{c}  \\
\texteltid{a4} & \texteltid{t4} & \texteltid{a6} & \texteltid{t5} & \texteltid{a3} & \texteltid{t2} & \texteltid{a2} & \texteltid{t3} & \texteltid{a4} & \texteltid{li4} & \texteltid{c2} \\
\end{pathbinding}

\noindent Note that variables with different subscript or superscript (e.g., $\varb{1}$ and $\varb{2}$) are not joined on.

Restrictors are also checked at this point. For example,  $\pi_{8,\gqltext{City}}$ has no match. Indeed, a path binding 
computed by the above join would 
use the loop (\texteltid{t4},\texteltid{t5},\texteltid{t2},\texteltid{t3}) twice, and thus would not be a trail.

In the end,~$\pi_{n,\ell}$ never has any matches unless~$n=4$ (as presented above) or~$n=7$.
The patterns $\pi_{4,\gqltext{City}}$, $\pi_{4,\gqltext{Country}}$, $\pi_{7,\gqltext{City}}$ and $\pi_{7,\gqltext{Country}}$ each have one match given below.
\begin{pathbinding}
  \gqltext{a} 
  & $\varb{1}$ & $\varnode[1]{2}$ 
  & $\varb{2}$ & $\varnode[2]{2}$
  & $\varb{3}$ & $\varnode[3]{2}$
  & $\varb{4}$ & \gqltext{a} & $\varedge{1}$ & \gqltext{c}  \\
\texteltid{a4} & \texteltid{t4} & \texteltid{a6} & \texteltid{t5} & \texteltid{a3} & \texteltid{t2} & \texteltid{a2} & \texteltid{t3} & \texteltid{a4} & \texteltid{li4} & \texteltid{c2} \\
\end{pathbinding}
\begin{pathbinding}
  \gqltext{a} 
  & $\varb{1}$ & $\varnode[1]{2}$ 
  & $\varb{2}$ & $\varnode[2]{2}$
  & $\varb{3}$ & $\varnode[3]{2}$
  & $\varb{4}$ & \gqltext{a} & $\varedge{2}$ & \gqltext{c}  \\
  \texteltid{a4} & \texteltid{t4} & \texteltid{a6} & \texteltid{t5} & \texteltid{a3} & \texteltid{t2} & \texteltid{a2} & \texteltid{t3} & \texteltid{a4} & \texteltid{li4} & \texteltid{c2} \\
\end{pathbinding}
\begin{pathbinding}
  \gqltext{a} 
  & $\varb{1}$ & $\varnode[1]{2}$ 
  & $\varb{2}$ & $\varnode[2]{2}$
  & $\varb{3}$ & $\varnode[3]{2}$
  & $\varb{4}$ & $\varnode[4]{2}$ 
  & $\varb{5}$ & $\varnode[5]{2}$
  & $\varb{6}$ & $\varnode[6]{2}$
  & $\varb{7}$ & \gqltext{a} & $\varedge{1}$ & \gqltext{c}  \\
\texteltid{a4} & \texteltid{t4} & \texteltid{a6} & \texteltid{t5} & \texteltid{a3} & \texteltid{t7} & \texteltid{a5} & \texteltid{t8} & \texteltid{a1} & \texteltid{t1} & \texteltid{a3} & \texteltid{t2} & \texteltid{a2} & \texteltid{t3} & \texteltid{a4} & \texteltid{li4} & \texteltid{c2} \\
\end{pathbinding}
\begin{pathbinding}
  \gqltext{a} 
  & $\varb{1}$ & $\varnode[1]{2}$ 
  & $\varb{2}$ & $\varnode[2]{2}$
  & $\varb{3}$ & $\varnode[3]{2}$
  & $\varb{4}$ & $\varnode[4]{2}$ 
  & $\varb{5}$ & $\varnode[5]{2}$
  & $\varb{6}$ & $\varnode[6]{2}$
  & $\varb{7}$ & \gqltext{a} & $\varedge{2}$ & \gqltext{c}  \\
   \texteltid{a4} & \texteltid{t4} & \texteltid{a6} & \texteltid{t5} & \texteltid{a3} & \texteltid{t7} & \texteltid{a5} & \texteltid{t8} & \texteltid{a1} & \texteltid{t1} & \texteltid{a3} & \texteltid{t2} & \texteltid{a2} & \texteltid{t3} & \texteltid{a4} & \texteltid{li4} & \texteltid{c2} \\
\end{pathbinding}

The latter two path bindings are trails, but are not acyclic, since the node \texteltid{a3} appears twice. Thus, they would 
have been filtered out if we had used the restrictor \gqlkw{ACYCLIC} instead of \gqlkw{TRAIL}.

\subsection{Reduction and Deduplication}
\label{s:reduction}
Reduction of path bindings strips variables from their annotations (subscripts and superscripts). In particular, it 
merges together all variables introduced in anonymous element patterns.
The result of reduction in our example is show below:

\begin{pathbinding}
  \gqltext{a} 
  & $\varb{}$ & $\rvarnode[]{}$ 
  & $\varb{}$ & $\rvarnode[]{}$
  & $\varb{}$ & $\rvarnode[]{}$
  & $\varb{}$ & \gqltext{a} & $\rvaredge{}$ & \gqltext{c}  \\
\texteltid{a4} & \texteltid{t4} & \texteltid{a6} & \texteltid{t5} & \texteltid{a3} & \texteltid{t2} & \texteltid{a2} & \texteltid{t3} & \texteltid{a4} & \texteltid{li4} & \texteltid{c2} \\
\end{pathbinding}
\begin{pathbinding}
  \gqltext{a} 
  & $\varb{}$ & $\rvarnode[]{}$ 
  & $\varb{}$ & $\rvarnode[]{}$
  & $\varb{}$ & $\rvarnode[]{}$
  & $\varb{}$ & \gqltext{a} & $\rvaredge{}$ & \gqltext{c}  \\
   \texteltid{a4} & \texteltid{t4} & \texteltid{a6} & \texteltid{t5} & \texteltid{a3} & \texteltid{t2} & \texteltid{a2} & \texteltid{t3} & \texteltid{a4} & \texteltid{li4} & \texteltid{c2} \\
\end{pathbinding}
\begin{pathbinding}
  \gqltext{a} 
  & $\varb{}$ & $\rvarnode[]{}$ 
  & $\varb{}$ & $\rvarnode[]{}$
  & $\varb{}$ & $\rvarnode[]{}$
  & $\varb{}$ & $\rvarnode[]{}$ 
  & $\varb{}$ & $\rvarnode[]{}$
  & $\varb{}$ & $\rvarnode[]{}$
  & $\varb{}$ & \gqltext{a} & $\rvaredge{}$ & \gqltext{c}  \\
   \texteltid{a4} & \texteltid{t4} & \texteltid{a6} & \texteltid{t5} & \texteltid{a3} & \texteltid{t7} & \texteltid{a5} & \texteltid{t8} & \texteltid{a1} & \texteltid{t1} & \texteltid{a3} & \texteltid{t2} & \texteltid{a2} & \texteltid{t3} & \texteltid{a4} & \texteltid{li4} & \texteltid{c2} \\
\end{pathbinding}
\begin{pathbinding}
  \gqltext{a} 
  & $\varb{}$ & $\rvarnode[]{}$ 
  & $\varb{}$ & $\rvarnode[]{}$
  & $\varb{}$ & $\rvarnode[]{}$
  & $\varb{}$ & $\rvarnode[]{}$ 
  & $\varb{}$ & $\rvarnode[]{}$
  & $\varb{}$ & $\rvarnode[]{}$
  & $\varb{}$ & \gqltext{a} & $\rvaredge{}$ & \gqltext{c}  \\
   \texteltid{a4} & \texteltid{t4} & \texteltid{a6} & \texteltid{t5} & \texteltid{a3} & \texteltid{t7} & \texteltid{a5} & \texteltid{t8} & \texteltid{a1} & \texteltid{t1} & \texteltid{a3} & \texteltid{t2} & \texteltid{a2} & \texteltid{t3} & \texteltid{a4} & \texteltid{li4} & \texteltid{c2} \\
\end{pathbinding}

\noindent All reduced path bindings are then collected together in a set.
That is, at this point, \emph{deduplication} occurs. If two different rigid patterns yield the same reduced path binding, then a single copy of the path binding is kept. 
Thus, the final result has only two distinct reduced path bindings.

\begin{pathbinding}
  \gqltext{a} 
  & $\varb{}$ & $\rvarnode[]{}$ 
  & $\varb{}$ & $\rvarnode[]{}$
  & $\varb{}$ & $\rvarnode[]{}$
  & $\varb{}$ & \gqltext{a} & $\rvaredge{}$ & \gqltext{c}  \\
   \texteltid{a4} & \texteltid{t4} & \texteltid{a6} & \texteltid{t5} & \texteltid{a3} & \texteltid{t2} & \texteltid{a2} & \texteltid{t3} & \texteltid{a4} & \texteltid{li4} & \texteltid{c2} \\
\end{pathbinding}
\begin{pathbinding}
  \gqltext{a} 
  & $\varb{}$ & $\rvarnode[]{}$ 
  & $\varb{}$ & $\rvarnode[]{}$
  & $\varb{}$ & $\rvarnode[]{}$
  & $\varb{}$ & $\rvarnode[]{}$ 
  & $\varb{}$ & $\rvarnode[]{}$
  & $\varb{}$ & $\rvarnode[]{}$
  & $\varb{}$ & \gqltext{a} & $\rvaredge{}$ & \gqltext{c}  \\
   \texteltid{a4} & \texteltid{t4} & \texteltid{a6} & \texteltid{t5} & \texteltid{a3} & \texteltid{t7} & \texteltid{a5} & \texteltid{t8} & \texteltid{a1} & \texteltid{t1} & \texteltid{a3} & \texteltid{t2} & \texteltid{a2} & \texteltid{t3} & \texteltid{a4} & \texteltid{li4} & \texteltid{c2} \\
\end{pathbinding}

\noindent The consequence of this deduplication is that our running query is 
equivalent to:

\begin{gql}
MATCH TRAIL
  (a WHERE a.owner='Jay') [-[b:Transfer WHERE b.amount>5M]->]+
  (a)-[:isLocatedIn]->(c:City|Country)
\end{gql}

\noindent Hence, a user does can turn a disjunction in the label expression into a path pattern union.

\bigskip

Having finished this detailed example, we now look at 
some other features of the pattern matching algorithm.

\paragraph{Using selectors}
After deduplication, any selectors (if present) would be applied.
Assume that in our running example we replaced the restrictor \gqlkw{TRAIL}
with the selector \gqlkw{ALL}~\gqlkw{SHORTEST}: 

\begin{gql}
MATCH ALL SHORTEST
  (a WHERE a.owner='Jay') [-[b:Transfer WHERE b.amount>5M]->]+
  (a) [-[:isLocatedIn]->(c:City) | -[:isLocatedIn]->(c:Country)]
\end{gql}

\noindent
While 
the number of reduced path bindings would be infinite (since \gqltext{Transfer} loops may be taken arbitrarily many times without the trail restriction), 
the selector would keep the shortest reduced binding path for each pair of endpoints (in our case, 
\texteltid{a4} and \texteltid{c2}), thus returning

\begin{pathbinding}
  \gqltext{a} 
  & $\varb{}$ & $\rvarnode[]{}$ 
  & $\varb{}$ & $\rvarnode[]{}$
  & $\varb{}$ & $\rvarnode[]{}$
  & $\varb{}$ & \gqltext{a} & $\rvaredge{}$ & \gqltext{c}  \\
   \texteltid{a4} & \texteltid{t4} & \texteltid{a6} & \texteltid{t5} & \texteltid{a3} & \texteltid{t2} & \texteltid{a2} & \texteltid{t3} & \texteltid{a4} & \texteltid{li4} & \texteltid{c2} \\
\end{pathbinding}

\paragraph{Multiple patterns}
When more than one pattern, separated by commas, appear after \gqlkw{MATCH}, 
each path pattern is 
solved separately, 
resulting in 
finitely many solutions to 
each path pattern. At this point the cross product
of the sets of path bindings 
is formed, and then filtered on the basis of 
implicit equi-joins on the global singleton variables
and the final \gqlkw{WHERE} clause.

\paragraph{Path pattern union vs multiset alternation}
The consequence of using deduplication is that our running query with path pattern union is 
equivalent to a query where the last edge pattern is replaced by 
\lstinline{(a)-[:isLocatedIn]->(c:City|Country)}.

To avoid deduplication and to maintain four reduced path bindings in the output, one could use multiset alternation instead, replacing the pattern by

\begin{gql}
 (a) [-[:isLocatedIn]->(c:City) |+| -[:isLocatedIn]->(c:Country)]
\end{gql}

\subsection{Query Outputs}\label{gql-output-sec}

\begin{figure}
    \resizebox{\linewidth}{!}{
    \begin{tikzpicture}
        \tikzstyle{every state}=[draw,rectangle,rounded corners,font=\ttfamily]

        \node[state] (graph_pattern) at (-4,1) {
          \begin{tabular}{l}
            \small Graph pattern\\
          \end{tabular}
        };

        \node[state] (graph_db) at (-4,-1) {
          \begin{tabular}{l}
            \small Graph DB\\
          \end{tabular}
        };

        \node[state] (gpml_processor) at (-2,0) {
          \begin{tabular}{l}
            \small GPML processor\\
          \end{tabular}
        };

        \node[state] (mp_bindings) at (1.5,0) {
          \begin{tabular}{l}
            \small output\\
          \end{tabular}
        };

        \node[state] (pgq) at (3,1) {
          \begin{tabular}{l}
            \small SQL/PGQ\\
          \end{tabular}
        };
        
        \node[state] (pgq_table) at (5,1) {
          \begin{tabular}{l}
            \small table\\
          \end{tabular}
        };

        \node[state] (gql) at (3,-1) {
          \begin{tabular}{l}
            \small GQL\\
          \end{tabular}
        };

        \node[state] (gql_table) at (5, 0) {
          \begin{tabular}{l}
            \small table\\
          \end{tabular}
        };

        \node[state] (graph_view) at (5,-1) {
          \begin{tabular}{l}
            \small graph view\\
          \end{tabular}
        };

        \node[state] (new_graph) at (5,-2) {
          \begin{tabular}{l}
            \small new graph\\
          \end{tabular}
        };

    \begin{pgfonlayer}{background}        
        \path[-latex]
        (graph_pattern) edge[bend left] (gpml_processor)
        (graph_db) edge[bend right] (gpml_processor)
        (gpml_processor) edge[] (mp_bindings)
        (mp_bindings) edge[bend left] (pgq)
        (pgq) edge[] (pgq_table)
        (mp_bindings) edge[bend right] (gql)
        (gql) edge[bend left] (gql_table)
        (gql) edge[] (graph_view)
        (gql) edge[bend right] (new_graph)
        ;

    \end{pgfonlayer}
    \end{tikzpicture}
    }
\caption{Conceptual diagram of GPML, SQL/PGQ and GQL}
\label{fig.gpmldiagram}
\end{figure}
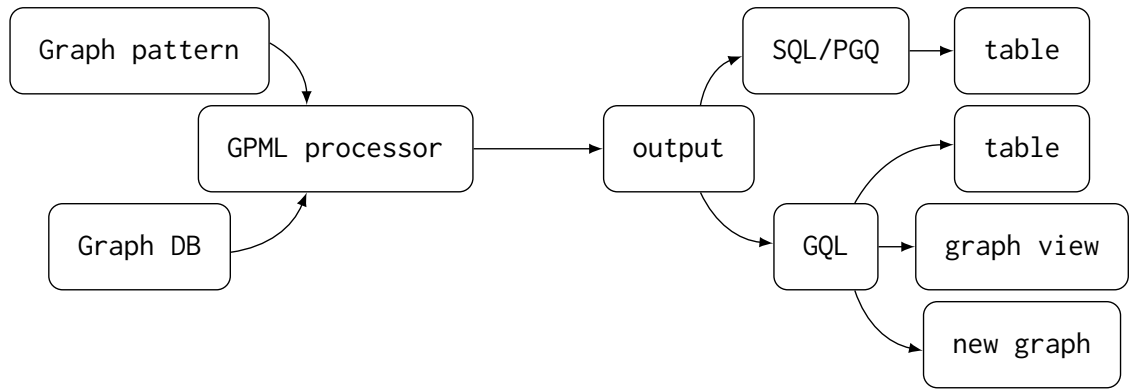

How should the result of pattern matching be represented to produce the output of a query? We have seen that executing a GPML statement results in a set of path bindings. Presenting this to the user depends on the host language, SQL/PGQ or GQL. 
Figure \ref{fig.gpmldiagram} shows 
the relationship between GPML and its two 
host languages. 

 The GPML processor is a software component within
    an implementation of either SQL/PGQ or GQL.  The two inputs to the GPML processor are the 
    graph pattern and the graph database.
     The output of the GPML processor is 
     consumed by the host 
     to produce the final
    output requested by the user. For SQL/PGQ, it will be a table, obtained from the computed reduced path bindings. 
    
    For GQL, the output could be more varied, including a 
    graph view, or new graph. Indeed, each path binding defines a subgraph of the input graph given by its nodes and edges, together with annotations, given by variables assigned to them in the path binding. This opens up more possibilities for structuring query outputs. While in the initial release of the GQL standard, outputs will be in line with those of SQL/PGQ, it is anticipated that in the future more advanced options will be added.

\OMIT{ 
\subsection{Comma-Separated List of Path Patterns}
When there is more than one path pattern, they are 
solved separately, including the evaluation of 
selectors, if any. Note that after evaluation of 
selectors, there are finitely many solutions to 
each path pattern. At this point the cross product
of the sets of path bindings 
is formed, and then filtered on the basis of 
implicit equijoins on the global singleton variables
and the final WHERE clause.
}

\section{Looking forward}\label{sec:lf}
In this section we outline the ongoing work on the development of the SQL/PGQ and GQL standards and list several research problems that have arisen in the process of designing the GPML.

\subsection{Standards Process: Steps and Timing}
The SQL/PGQ and GQL standards are being developed in the international standards committee ISO/IEC JTC1 SC32 WG3 "Database Languages"
with input from various national bodies. In particular, the US  committee INCITS DM32 "Data Management and Interchange" and DM32's SQL/PGQ and GQL expert groups review all significant US change proposals before they are considered by WG3.

The ISO/IEC JTC1 process has a number of steps with ballots to transition between the stages. The high-level overview is:
\begin{itemize}
    \item Initial effort -- develop and expand the draft;
    \item Committee Draft (CD) Ballot -- 12 weeks;
    \item Draft International Standard (DIS) Ballot -- 20 weeks;
    \item International Standard (IS) published. 
\end{itemize}
After each ballot, time is needed to resolve the comments submitted.

The current schedule for the progression of the SQL/PGQ and GQL standards is shown in Figure \ref{fig.timeline}.\footnote{The schedule depends on work that has not been completed and so could change.} By the time the DIS ballot starts, the technical specification is fairly stable.
Since GPML is the same for GQL and SQL/PGQ, GQL GPML will be fairly stable when SQL/PGQ begins DIS ballot.
\begin{figure}
\begin{tabular}{lcc}
\toprule
 \textbf{Date} & \textbf{SQL/PGQ} & \textbf{GQL} \\ 
\midrule
 2017 & Work started & \\
 2018 & & Work started \\
 2021-02-07 & CD Ballot End &  \\  
 2022-02-20 &  & CD Ballot End \\
 2022-12-04 & DIS Ballot End & \\
 2023-01-30 & Final Text to ISO & \\
 2023-03-13 & \textbf{SQL/PGQ IS Published} & \\
  2023-05-21 & & DIS Ballot End \\
 2023-07-30 & & Final Text to ISO \\
 2023-09-10 & & \textbf{GQL IS Published} \\
\bottomrule
\end{tabular}
\caption{SQL/PGQ and GQL Timeline}
\label{fig.timeline}
\end{figure}
As SC32 WG3 makes progress on the drafts, it accumulates \emph{Language Opportunities} (LOs). LOs are capabilities that are potentially useful, but are not yet ready for the current versions of the draft standards. Below we provide a sample of LOs pertaining to GPML:
\begin{itemize}
    \item Constraining a graph pattern through the introduction of isomorphic match modes: for example, an \emph{edge-isomorphic} match requires all edges matched across all constituent path patterns in the graph pattern to differ from each other. 
    \item Queries on multiple graphs in a single concatenated \gqlkw{MATCH}.
    \item 
      Path macros for multiple use in a query.
    \item Outputting the interleaving of bindings in nested quantifiers, such as \gqlinline![[(p)->(q)]* ->(r)]*!.
    \item Cheapest path search, by adding weights to edges.
    \item Exporting a graph element or path binding to JSON.
\end{itemize}

GQL also has LOs that go beyond those common with SQL/PGQ. Examples 
include property graph keys and constraints \cite{pg-keys}, 
system versioned graphs, 
and stored queries, procedures, and functions.
As discussed in Section \ref{gql-output-sec}, formats such as JSON could potentially be used for
returning a raw multi-path binding.

\subsection{Research Questions}
There are many open questions related to GPML. Given the novelty of the language, and its goal as the future standard of querying property graphs, it is important to full understand the expressiveness and  complexity of its various fragments, in the spirit of many decades of research tradition in database query languages \cite{ABLMP21}. 
    
Some of the most intriguing questions concern processing unbounded paths. For instance, the innocently looking query below may not terminate.

\begin{gql}
MATCH (x)-[e]->*(y)
WHERE AVG(e.a)<1
KEEP ANY SHORTEST
\end{gql}

\noindent With a sufficiently rich pattern language, termination will become an undecidable problem. A natural question is then whether there are interesting classes of predicates on aggregates of group variables for which termination can be guaranteed, or efficiently checked at compile time.

There are many questions related to the implementation of GPML. How does one solve efficiently shortest path queries with arbitrary regular expressions, not just \gqlinline{->*} as in Dijkstra's  algorithm? Can we handle more complex optimization problems, such as maximizing an objective function subject to an upper bound on the length or cost of the path (e.g., ``What is the most scenic route to the airport in at most 2 hours?''). 

Another direction is to consider fully recursive graph patterns, permitting multiple self-references, not just a single one like in the \gqlinline{*} operator. Such patterns might be used to search for trees and other structures more complex than paths. Is there intuitive syntax to express such patterns? What real-world problems might they address? What is the cost of adding them to GPML?

Yet another line of future works concerns extended capabilities and different uses of graph patterns. One of them concerns extending the language to capture the temporal aspect of data \cite{MS-dbpl17,RGFTR21}. A different line of work is to understand the underpinnings of languages defined by paths in property graphs, which are more than traditional regular languages over a finite alphabet, as they include property values coming from potentially infinite domains. Simple models for this based on data words \cite{BDMSS11} have already been studied in \cite{gxpath}, but they disregarded operations on data, such as arithmetic or string operations.  In an opposite direction, we may look at restrictions that will be useful in different applications, such as for example updating property graphs, which tends to be problematic \cite{cypher-updates} with a full power of pattern matching.

\section{Related Work}\label{sec:related}

GPML is the culmination of a long evolution of graph query languages, including contributions in research, standards and practice, too rich to survey comprehensively here. 
We  discuss related work only in broad strokes and refer the reader to existing surveys \cite{Woo,surveyChile}.

GPML's graph patterns extend the celebrated language of Conjunctive Regular Path Queries (CRPQ) introduced in~\cite{CruzMW87} and studied in a plethora of follow-up work (e.g.~\cite{BarceloLLW-tods12,BarceloLR-jacm14,CGLV03,CalvaneseGLV99,CGLVkr,FigueiraGKMNT-kr20}). 
CRPQs consist of variables (typically binding only to individual vertices and edges), path patterns specified by regular expressions over the edge labels, and comma-separated lists thereof. 
In their first incarnation, CRPQs operated over property-less graphs, but subsequent work considered graphs with data annotations (e.g.~\cite{gxpath}), which are ancestors of today's property graphs.
Notable early graph query languages derived from CRPQs include G~\cite{CruzMW87}, StruQL~\cite{struql}, 
Lorel~\cite{lorel} and WebSQL~\cite{MendelzonMM96}.
Syntactically, GPML graph patterns extend CRPQs by introducing group variables and variables binding to entire paths. 
The latter allows GPML to treat paths as first-class citizens, as advocated in the G-CORE proposal~\cite{gcore}.
Semantically, graph patterns extend CRPQs by supporting a finer-grained notion of match, which permits binding variables to paths, using multiplicity-sensitive aggregation such as sum, count, and average, and restricting the number of possible returned paths. These features are original contributions of GPML.

The \emph{simple path} restrictor was already present in the first paper that introduced CRPQs \cite{CruzMW87}. 
However, after it was discovered that the evaluation problem for (C)RPQs with simple path and trail restrictors has a high computational complexity~\cite{MendelzonW95}, they were put on hold by the research community. They were resurrected by practical query languages (e.g., Cypher \cite{Cypher} and early versions of SPARQL 1.1.~\cite{sparql11}), which generated another wave of fundamental research \cite{BaganBG13,MartensT18,MartensNT20} that sheds a clearer light on the relationship between queries that appear in real-life and the high complexity of simple paths and trails in the worst case.

GPML patterns share many features with pattern standards developed for different data models. 
\emph{XML documents}~\cite{xml} correspond to tree-shaped, vertex-labeled graphs. 
The XPath~\cite{xpath} World-Wide Web Consortium (W3C) standard describes patterns that roughly correspond to GPML path patterns that can be combined into tree-shaped structures; the use of variables is limited.
The XQuery~\cite{xquery} standard supports comma-separated lists of XPath patterns, stitched together with node variables. 
Like in GPML, XQuery matches are discriminated by the path bindings they correspond to, but the limited use of quantifiers and the tree shape of the graph avoid the query non-termination problem. 
\emph{Knowledge Graphs}~\cite{kg2021} corresponding to the \emph{RDF} standard~\cite{rdf}, also admit pattern-based querying, for instance using the SPARQL W3C standard query language~\cite{sparql11}. 
Patterns consist once more of variables (binding to the RDF counterpart of nodes and edges) and regular path expressions. To deal with cycles, and potentially infinite number of matching paths~\cite{LM13,ACP12}, SPARQL deploys the ``endpoint semantics'' of paths, and will only return start/end point of a path, instead of the matching path itself.


\medskip
\noindent
{\bf ACKNOWLEDGMENTS} \ \ The work of the FSWG is supported by grants from Neo4j held at the University of Edinburgh and ENS-Paris. The academic group also gratefully acknowledges support of the following research grants: EPSRC grants N023056 and S003800 (Libkin); DFG grants 369116833 and 431183758 (Martens); NCN grant 2018/30/E/ST6/00042 (Murlak); ANID – Millennium Science Initiative Program – Code ICN17\_002 (Vrgo\v{c}).

\bibliographystyle{ACM-Reference-Format}
\bibliography{bibfile}

\end{document}